\documentclass[manuscript,screen,nonacm]{acmart}
\usepackage{color, colortbl, xcolor}
\usepackage{url}
\usepackage{subcaption}
\usepackage{textcomp}
\usepackage{soul}
\usepackage{multirow}
\usepackage{enumitem}
\usepackage{mathtools}
\usepackage{siunitx}
\usepackage{array}
\usepackage{colortbl}
\usepackage{hhline}
\usepackage{adjustbox}
\usepackage{makecell}
\usepackage{placeins}

\usepackage{booktabs} 
\usepackage{array}
\usepackage{xcolor}
\usepackage{tabularx}
\usepackage{placeins}
\usepackage{makecell}

\usepackage[table]{xcolor}
\usepackage{booktabs}
\usepackage{makecell}

\definecolor{sigblue}{RGB}{66, 133, 180}
\definecolor{sigorange}{RGB}{230, 126, 34}

\newcommand{\posone}[1]{\cellcolor{sigorange!12}#1}
\newcommand{\postwo}[1]{\cellcolor{sigorange!22}#1}
\newcommand{\posthree}[1]{\cellcolor{sigorange!34}#1}

\newcommand{\negone}[1]{\cellcolor{sigblue!12}#1}
\newcommand{\negtwo}[1]{\cellcolor{sigblue!22}#1}
\newcommand{\negthree}[1]{\cellcolor{sigblue!34}#1}

\newcommand*{\rowstyle}[1]{
  \gdef\@rowstyle{#1}%
  \@rowstyle\ignorespaces%
}

\newcolumntype{=}{
  >{\gdef\@rowstyle{}}%
}

\newcolumntype{+}{
  >{\@rowstyle}%
}

\usepackage{arydshln}
\definecolor{linkColor}{RGB}{6,125,233}
\definecolor{green}{rgb}{0.0, 0.65, 0.31}
\definecolor{bleudefrance}{rgb}{0.19, 0.55, 0.91}
\definecolor{ceruleanblue}{rgb}{0.16, 0.32, 0.75}
\definecolor{grey}{HTML}{969696}
\definecolor{violet}{HTML}{756bb1}
\definecolor{dgrey}{HTML}{01665e}
\definecolor{lgrey}{HTML}{5ab4ac}
\definecolor{dgreen}{HTML}{005a32}
\definecolor{purple}{HTML}{54278f}

\definecolor{editCol}{HTML}{000000}
\definecolor{maskCol}{HTML}{c51b7d}
\definecolor{lrColor}{HTML}{8856a7}
\definecolor{trColor}{HTML}{d01c8b}
\definecolor{ctColor}{HTML}{4dac26}
\definecolor{brickred}{HTML}{f03b20}
\definecolor{improveCol}{HTML}{253494}
\definecolor{worsenCol}{HTML}{d7191c}
\definecolor{DarkBlue}{HTML}{00008B}
\definecolor{mscolor}{HTML}{01665e}
\definecolor{nmscolor}{HTML}{bf812d}
\definecolor{lgreen}{HTML}{ccece6}
\definecolor{dolive}{HTML}{308014}
\definecolor{dpink}{HTML}{CD1076}
\definecolor{soothinggreen}{HTML}{4dac26}
\definecolor{darkred}{HTML}{8B0000}

\colorlet{tablerowcolor4}{gray!50} 

\newcommand*{\textlabel}[2]{%
  \edef\@currentlabel{#1}
  \phantomsection
  #1\label{#2}
}

\colorlet{tableheadcolor}{gray!25} 
\colorlet{tablerowcolor}{gray!10} 
\colorlet{tablerowcolor2}{gray!45} 
\colorlet{tablerowcolor3}{gray!25} 

\newcolumntype{a}{>{\columncolor{tablerowcolor}}r}
\definecolor{aicolor}{HTML}{018571}
\definecolor{occolor}{HTML}{ff7799}

\definecolor{aicolor}{HTML}{fc8d62}
\definecolor{occolor}{HTML}{253494}

\newif{\ifhidecomments}
 \hidecommentsfalse 
\ifhidecomments
    \newcommand{\jenny}[1]{}
    \newcommand{\koustuv}[1]{}
\else
    \newcommand{\jenny}[1]{\textbf{\small\sffamily{\textcolor{dgreen}{[#1 -- Jenny]}}}}
    \newcommand{\koustuv}[1]{\textbf{\small\sffamily{\textcolor{dpink}{[#1 -- Koustuv]}}}}
  \fi

\renewcommand{\textrightarrow}{$\rightarrow$}

\colorlet{tableheadcolor}{gray!25} 

\definecolor{neutralCol}{HTML}{dd1c77}
\definecolor{neutralGreen}{HTML}{31a354}
\definecolor{NewBlue}{HTML}{1879ba}
\definecolor{bleudefrance}{rgb}{0.19, 0.55, 0.91}  
\definecolor{AfTrColor}{HTML}{0868ac}  
\definecolor{BfTrColor}{HTML}{a8ddb5}  

\definecolor{AfCtColor}{HTML}{b10026}  
\definecolor{BfCtColor}{HTML}{fd8d3c}

\graphicspath{ {figures/} }

\newcolumntype{C}[1]{>{\centering\arraybackslash}p{#1}}

\AtBeginDocument{%
  \providecommand\BibTeX{{%
    \normalfont B\kern-0.5em{\scshape i\kern-0.25em b}\kern-0.8em\TeX}}}

\begin{document}



\title[Trustworthy FinAInce: Unpacking How AI-Mediated Financial Advice is Judged]{Trustworthy FinAInce: Unpacking How AI-Mediated Financial Advice is Judged}

\author{Aryan Ramchandra Kapadia}
\orcid{0009-0007-9134-603X}
\email{kapadia8@illinois.edu}
\affiliation{%
  \institution{University of Illinois Urbana-Champaign}
  \city{Urbana}
  \state{IL}
  \country{USA}
}

\author{Eshwar Chandrasekharan}
\orcid{0000-0002-7473-1418}
\email{eshwar@illinois.edu}
\affiliation{%
  \institution{University of Illinois Urbana-Champaign}
  \city{Urbana}
  \state{IL}
  \country{USA}
}

\author{Koustuv Saha}
\orcid{0000-0002-8872-2934}
\email{ksaha2@illinois.edu}
\affiliation{%
  \institution{University of Illinois Urbana-Champaign}
  \city{Urbana}
  \state{IL}
  \country{USA}
}


\renewcommand{\shortauthors}{Kapadia et al.}

\begin{abstract}

As generative AI is increasingly used as a source of personal financial guidance, understanding how people appraise such advice is important for supporting appropriate reliance. We conducted a randomized vignette experiment with 285 U.S. adults across eight financial decisions, independently varying three advice styles---\textit{AI}, \textit{expert}, and \textit{online community}---and displayed source labels while holding the underlying recommendation consistent. Advice style most strongly shaped message and safety appraisals, Expert labels selectively increased perceived source knowledge, and decision context primarily shaped risk and safety appraisals. These appraisals were associated with downstream judgments, with models explaining 69.2\% of overall quality, 75.9\% of trust, and 82.9\% of intended reliance. Expert-style advice also remained most preferred when shown without source labels. Our findings have implications for understanding financial advice evaluation, distinguishing the roles of advice style and source labels, and designing financial AI that supports grounded evaluation rather than simply maximizing trust.

\end{abstract}




\maketitle


\section{Introduction}



Generative AI is rapidly becoming a source of everyday financial guidance,
from budgeting and saving to investing~\cite{choukhmaneAIFinancialAdvice2026,pakHowIndividualsUse2026}.
General-purpose conversational AI systems such as \textit{ChatGPT}, \textit{Gemini}, and \textit{Claude}, as well as finance-oriented tools such as \textit{SoFi Coach}~\cite{WhatSoFiLearned}, make financial guidance available on demand, often at little to no direct cost, while conversational interfaces enable context disclosure and iterative refinement~\cite{xiao2025ai,pakHowIndividualsUse2026}.
This accessibility is already translating into substantial use: in a recent survey of 4,000 U.S. consumers, 51\% reported using AI for financial advice or information~\cite{MoreUSConsumers2026}.
Yet easier access does not guarantee reliability.
LLMs can still provide unreliable or insufficiently contextualized financial
guidance, alongside broader concerns around privacy, transparency, and
accountability~\cite{jiSurveyHallucinationNatural2023,
huangSurveyHallucinationLarge2025,oecdArtificialIntelligencePersonal2026,
heanCanAIHelp2025}. Moreover, their performance can deteriorate as financial decisions become more complex and context-dependent
~\cite{heanCanAIHelp2025,choukhmaneAIFinancialAdvice2026}.
This is particularly consequential in personal finance, where plausible but imperfect recommendations can cause financial harm, compounded by overreliance~\cite{bucincaTrustThinkCognitive2021,klingbeilTrustRelianceAI2024}, growing AI dependency~\cite{gohGenerativeArtificialIntelligence2025}, and cognitive offloading of financial judgment~\cite{riskoCognitiveOffloading2016,zhuNotAllCognitive2026}. Understanding how users evaluate such advice---and how those appraisals relate to trust and intended reliance---is therefore important for designing financial AI that supports appropriately grounded decision-making.

However, the process of advice evaluation is much more complex than assessing substantive content alone.~\cite{bonaccioAdviceTakingDecisionmaking2006}  Prior work shows that message-level characteristics such as tone, reasoning, and presentation can shape responses to advice~\cite{chengHumanVsAI2022a,metzgerEmpoweringCalibratedDistrust2024,
takayanagiAreGenerativeAI2025,sunBeFriendlyNot2026}, while explicit
\emph{source labels} can activate expectations about expertise, credibility,
and authority
~\cite{gallegosLabelingMessagesAIgenerated2026,
sunUnderstandingTrustHuman2026,sahebiAIPenaltyDisclosure2026}.
Advice evaluation may further vary with the \emph{decision context}
~\cite{bonaccioAdviceTakingDecisionmaking2006,
ioannouRoleExplainabilityAIDriven2026} and with the knowledge, experiences,
and predispositions of the person evaluating it
~\cite{allgoodEffectsPerceivedActual2016,grawitchWhoListensChatGPT2026}. Yet these influences have largely been examined separately or only in partial combinations, leaving limited understanding of how they jointly shape advice evaluation and where differences in \textit{trust} and intended \textit{reliance} originate.

Understanding how these influences shape advice evaluation becomes especially important in a multi-source financial advice ecosystem, where people may encounter guidance from AI systems, financial professionals, and online communities that differ in perceived expertise, credibility, and accountability. Yet financial-AI research has largely compared AI with professional advisors or examined AI advice in isolation, with comparatively little work examining online-community advice within the same evaluative framework
~\cite{kumtaCanYouSmell2026,yangMyAdvisorHer2026,
zarifisHowBuildTrust2024,takayanagiAreGenerativeAI2025,
thukralGeneratingInsightsFinancial2024}. These source perceptions may arise from both explicit and implicit cues: \emph{source labels} directly indicate who or what produced the advice, whereas \emph{advice style} conveys source-associated signals through tone, reasoning, and presentation. Distinguishing these cues matters because they have different implications for financial AI design and governance: advice style shapes how authority is communicated, whereas source labels make provenance explicit. Lastly, the decision context may influence how people weigh different cues when evaluating advice, with implications for subsequent \textit{trust} and \textit{reliance}~\cite{bonaccioAdviceTakingDecisionmaking2006, ioannouRoleExplainabilityAIDriven2026}. 
Motivated by these gaps, we ask the following research questions (RQs):

\paragraph{RQ1:} What factors shape people's appraisals of financial advice?

\paragraph{RQ2:} How are these appraisals associated with \textit{Overall Quality}, \textit{Trust}, and \textit{Intended Reliance}? \footnote{\textit{Overall Quality} refers to participants' overall evaluation of the advice; \textit{Trust} refers to whether they would trust the advice if they were in the protagonist's situation; and \textit {Intended Reliance} refers to
whether they would feel comfortable following the advice.}

To answer these RQs, we conducted a preregistered randomized vignette
experiment with 285 U.S. adults, yielding 1,140 repeated evaluations of
financial advice. We constructed eight personal-finance scenarios spanning a
$2\times2\times2$ decision-context taxonomy of \textit{stakes}, \textit{external uncertainty},
and \textit{verifiability}. For each scenario, we constructed three source-associated \emph{advice styles}---AI Financial Assistant, Certified Financial Planner, and Online Community---while keeping the underlying financial recommendation consistent across versions. We independently
manipulated \emph{source labels} by presenting advice with its corresponding
label, without a label, or with an incorrect label. Guided by the Information
Adoption Model (IAM) as an organizing lens for information evaluation and
adoption~\cite{sussmanInformationalInfluenceOrganizations2003}, we examined
advice style, source labels, decision context, and individual differences as
potential influences on four appraisals---\emph{Message Appraisal},
\emph{Risk Acknowledgment}, \emph{Safety Concern}, and \emph{Source
Knowledge}---and how these appraisals were associated with \emph{Overall
Quality}, \emph{Trust}, and \emph{Intended Reliance}. We additionally included
a comparative task in which participants ranked all three advice styles for a
previously unseen financial scenario.

Our results show that advice style, source labels, and decision context were associated with different aspects of financial advice evaluation. \emph{Advice style} showed broader and more consistent associations across appraisals, particularly with \emph{Message Appraisal} and \emph{Safety Concern}, whereas \emph{source-label} effects were more selective and most clearly reflected in perceived \emph{Source Knowledge}. \emph{Decision context} was associated primarily with risk- and safety-related appraisals. These appraisals were much more strongly associated with downstream outcomes than upstream factors (advice style, source labels, and decision context) alone: models incorporating the evaluation pathway explained 69.2\% of the variance in \emph{Overall Quality}, 75.9\% in \emph{Trust}, and 82.9\% in intended \emph{Reliance}, compared with 9.8--14.8\% for upstream factors alone. \emph{Overall Quality} was the strongest correlate of \emph{Trust}
($\beta=.607$), while \emph{Trust} was the strongest correlate of 
\emph{Intended Reliance} ($\beta=.594$). We also find that the effects of Advice style and source labels were largely additive rather than interactive, while individual differences produced more selective associations with how appraisal cues related to \emph{Trust} and
\emph{Intended Reliance}.

Overall, our work makes empirical, theoretical, and design contributions.
\textbf{Empirically}, we compare AI, Expert, and Online Community financial advice within a common experimental framework, disentangling \emph{advice style} from \emph{source labels} while examining how decision context and individual differences relate to advice evaluation.
\textbf{Theoretically}, we use information adoption as an organizing lens to show that \textit{Trust} and intended \textit{Reliance} are embedded within a broader evaluation process: different factors are associated with distinct appraisals, which are in turn strongly associated with \textit{Overall Quality}, \textit{Trust}, and intended \textit{Reliance}.
From a \textbf{design} perspective, we argue for an evaluation-first approach to financial AI: systems should help users assess the reasoning, risks, and provenance underlying advice, rather than simply making recommendations appear more trustworthy. 
Together, these contributions shift the design goal from maximizing \textit{Trust} toward supporting grounded evaluation and appropriate \textit{Reliance}.

\section{Related Work}

\subsection{Financial Advice in the Age of AI}

Recent advances in large language models have expanded their potential role in
personal financial decision support. Newer systems demonstrate growing
capabilities across financial analysis, investment-related tasks, portfolio
support, and personal financial guidance
~\cite{fairhurstHowMuchDoes2025,heanCanAIHelp2025,
niszczotaGPTHasBecome2023,kirtacSentimentTradingLarge2024,
fatourosCanLargeLanguage2025,schloskyChatGPTFinancialAdvisor2025}.
Their natural-language interfaces also make financial guidance highly
accessible, allowing users to describe their circumstances, ask follow-up
questions, and progressively refine recommendations
~\cite{xiao2025ai,pakHowIndividualsUse2026}. However, technical
capability alone does not guarantee reliable personalized advice. LLMs can
produce inaccurate information, operate with incomplete user context, and
raise privacy, transparency, and compliance concerns
~\cite{jiSurveyHallucinationNatural2023,
huangSurveyHallucinationLarge2025,oecdArtificialIntelligencePersonal2026};
their performance can also deteriorate as financial decisions become more
complex and context-dependent~\cite{heanCanAIHelp2025}. Similarly,
\citet{choukhmaneAIFinancialAdvice2026} find that LLM recommendations broadly
align with established financial principles while remaining less reliable for
subtler, context-sensitive aspects of household financial planning. These
limitations have motivated growing human-centered research on how people
perceive and use AI-generated financial advice, including work on advisor
preferences, trust, personalization, transparency, and interaction design
~\cite{merkleAIAppreciationFinancial2025,bosompimWhomWeTrust2026,
zarifisHowBuildTrust2024,takayanagiAreGenerativeAI2025}. We build on this shift beyond technical performance toward understanding how people perceive, evaluate, and ultimately use AI-generated financial advice.

Financial advice, however, is rarely encountered in isolation from other
sources. Recent nationally representative evidence shows that among the U.S. adults who sought financial guidance in the preceding year, 32\% consulted professional advisors; informal and digitally mediated sources included family members (35\%), friends (23\%), internet research (73\%), and news, media, or social media (26\%); and 18\% used AI tools such as ChatGPT or
Claude~\cite{incWhereAmericansCanadians2026}. At the same time, source use did not directly correspond to confidence, which was highest for professional advisors and substantially lower for AI systems
~\cite{incWhereAmericansCanadians2026, akanaTrustCredibilityComparinga}. Financial decision-making is also socially embedded: peer effects can shape
investment behavior~\cite{ouimetLearningCoworkersPeer2020}, online communities
such as Reddit influence how retail investors interpret financial information
and develop strategies~\cite{munsterRobinhoodRedditNews2024}, and Trust in
financial guidance can vary with both the advice source and users' surrounding
social networks~\cite{bosompimWhomWeTrust2026}. Yet research on AI-mediated
financial advice has largely examined AI relative to professional advisors
~\cite{merkleAIAppreciationFinancial2025,bosompimWhomWeTrust2026},
AI--human hybrid advisors~\cite{kumtaCanYouSmell2026,yangMyAdvisorHer2026},
or AI systems in isolation through factors such as personalization, preference-elicitation, agent personality, and human-like interaction
~\cite{takayanagiAreGenerativeAI2025,zarifisHowBuildTrust2024}. Online communities have received comparatively separate attention; for
example, \citet{thukralGeneratingInsightsFinancial2024} compare
LLM-generated and human responses to financial questions on Reddit. What
remains less understood is how people evaluate AI, professional, and
online-community advice within the same financial decision setting. We build
on this literature by examining these three prominent advice sources within a
common evaluation framework.

\subsection{Evaluating Financial Advice: Appraisals, Trust, and Reliance}

The Information Adoption Model (IAM) provides a useful organizing lens for
understanding how people evaluate and ultimately use external information.
At its core, IAM proposes that people engage in a sequence of cognitive
assessments when encountering information, including evaluations of
\textit{argument quality} and \textit{source credibility}, which shape
perceived \textit{information usefulness} and subsequent
\textit{information adoption}
~\cite{sussmanInformationalInfluenceOrganizations2003}. This perspective has
been widely applied in digital information environments
~\cite{vermaUnderstandingImpactEWOM2023,
erkanInfluenceEWOMSocial2016} and, more recently, to AI-mediated advice
~\cite{ioannouRoleExplainabilityAIDriven2026,
liHowUsersAdopt2025,gongWhenAlgorithmsSpeak2026,
camilleriAcceptanceUsageChatGPT2024}. In financial AI specifically, recent
work has further incorporated constructs such as perceived trust and risk into
this broader evaluation and adoption process
~\cite{ioannouRoleExplainabilityAIDriven2026}. In this study, we build on this
perspective by examining these cognitive assessments at
greater granularity in the context of financial advice. In addition to
message- and source-related appraisals, we make \textit{Risk Acknowledgment}
and \textit{Safety Concern} explicit because financial advice can appear clear
and credible while still failing to acknowledge uncertainty or exposing users
to potential harm.

These cognitive assessments ultimately matter because they shape whether advice
is trusted and acted upon. While IAM focuses on information adoption as the
downstream outcome of evaluation, human--AI research distinguishes
\textit{trust} from \textit{reliance}. Trust generally reflects a user's
attitude toward the competence, reliability, or trustworthiness of an advisor
or system, whereas reliance concerns whether its recommendation is incorporated
into a decision
~\cite{leeTrustAutomationDesigning2004,
raeesTrustRelianceMeasurement2026}. The two are related but not
interchangeable: users may report Trust without following a recommendation, or
rely on advice without correspondingly high reported trust
~\cite{raeesTrustRelianceMeasurement2026, raeesPeopleAppropriatelyRely2026, palWeKnowWhat2026}. Advice-taking research similarly emphasizes that receiving or valuing advice is distinct from the extent to
which it is incorporated into one's final judgment
~\cite{bonaccioAdviceTakingDecisionmaking2006}. This motivates us to treat \textit{Overall Quality}, \textit{Trust}, and
\textit{Intended Reliance} as distinct downstream responses to the appraisals users form.

Moreover, this framework is particularly relevant to financial decision-making contexts~\cite{ioannouRoleExplainabilityAIDriven2026}. Characteristics of 
decision context may shape which cues become salient and how people interpret them.
For example, the magnitude of potential consequences can affect how carefully
people approach a decision
~\cite{kahnExploratoryStudyChoice1995,slovicPerceptionRisk1987}, uncertainty
can increase the value of external guidance
~\cite{bonaccioAdviceTakingDecisionmaking2006}, and differences in
verifiability can affect how readily the quality of a recommendation can be
assessed against established principles
~\cite{laughlinSocialCombinationProcesses1980,
bonaccioAdviceTakingDecisionmaking2006}. Finance-specific work similarly shows
that decision risk can alter how information quality contributes to perceived
usefulness
~\cite{ioannouRoleExplainabilityAIDriven2026}. The same evaluative process may also vary across individuals. Financial
knowledge, risk tolerance, familiarity with a decision, and perceived expertise
can affect users' ability or confidence in evaluating external advice
~\cite{allgoodEffectsPerceivedActual2016,
grableFinancialRiskTolerance1999,
grawitchWhoListensChatGPT2026,
yanivAdviceTakingDecision2000}, while AI literacy, prior experience, and
broader orientations toward AI can shape expectations about AI-mediated advice
~\cite{alruwailiModelingPublicTrust2025,
huangInfluenceAILiteracy2024,
panAILiteracyTrust2025,
jessupMeasurementPropensityTrust2019}. Advice-taking and persuasion research similarly suggest that decision context and individual characteristics can shape how people weight different
evaluative cues when considering external advice
~\cite{bonaccioAdviceTakingDecisionmaking2006,
kammerSystematicReviewEmpirical2023,
chaikenHeuristicSystematicInformation1980,
pettyElaborationLikelihoodModel1986}. We therefore consider these factors as potential boundary conditions on this information-evaluation process, examining whether they shape the appraisals
users form and moderate the downstream associations of those appraisals with
\textit{Trust} and \textit{Intended Reliance}.

\subsection{Explicit and Implicit Cues in Financial Advice: Source Labels vs.\ Advice Style}

Advice is evaluated not only through its substantive content, but also through
cues about who or what produced it. Source information can activate
expectations about expertise, credibility, and competence
~\cite{chaikenHeuristicSystematicInformation1980,
pettyElaborationLikelihoodModel1986}, and human--AI studies show that responses
to otherwise comparable advice can differ depending on whether it is labeled
as coming from a human or algorithmic source
~\cite{vodrahalliHumansTrustAdvice2022,
loggAlgorithmAppreciationPeople2019}. Source identity, however, is not
communicated through labels alone. We use \textit{advice style} to refer to
message-level cues such as voice, tone, reasoning organization, experiential
grounding, expressions of authority, and recommendation framing. Professional
financial advice commonly emphasizes client circumstances, financial
principles, trade-offs, and longer-term consequences
~\cite{fpsbStandardsProfession,
macdonaldValuePersonalProfessional2023}, whereas online financial communities
more often incorporate clarification, multiple viewpoints, empathy, and
personal experience~\cite{thukralGeneratingInsightsFinancial2024}.
Comparisons of AI- and human-authored responses similarly identify differences
in linguistic and interactional patterns, including more structured and
neutral presentation in AI responses and greater conversational or
experiential grounding in human responses
~\cite{thukralGeneratingInsightsFinancial2024,
sahaAIVsHumans2026,sahaLinguisticComparisonAI2026}. 
Recent work similarly shows that generative AI can reproduce source-associated identity cues, including language suggestive of lived experience, without possessing the underlying experience those cues imply~\cite{goel2026synthetic}.
Such stylistic differences
can themselves affect reader preferences and perceptions
~\cite{zhouCommunicationStylesReader2026,
takayanagiAreGenerativeAI2025, wangMutualTheoryMind2021}. We therefore treat these patterns not as fixed
properties of AI, professional, or community sources, but as recognizable
source-associated communication profiles that may provide evaluative cues
independently of an explicit source label.

Distinguishing the effect of advice style from source labels matters because these mechanisms have distinct design implications for how trust and reliance are shaped. Prior work
faces a trade-off between experimental control and ecological validity.
Identical-text designs can isolate the effect of source labels but remove the
communication cues through which source identities are often conveyed in
practice
~\cite{vodrahalliHumansTrustAdvice2022,
loggAlgorithmAppreciationPeople2019}; naturalistic comparisons preserve those
cues but may simultaneously vary substantive content, reasoning, style, and
advice quality
~\cite{ayersComparingPhysicianArtificial2023,
kumarWhenAIGives2026}. This makes it difficult to determine whether observed
differences arise from \textit{how} advice is communicated or from
\textit{who} users believe produced it. This distinction also matters for AI governance, because professional presentation can be reproduced independently of the expertise, review, or
accountability that a source label may imply. We address this gap by standardizing
financial facts, numerical values, core reasoning, and recommendation direction
across AI-, Expert-, and Online Community-style advice, while independently
varying the source labels displayed alongside it. 
\section{Study Design and Data \label{section:data}}

\subsection{Study Overview}

\begin{figure}[htbp]
    \centering
    \includegraphics[
        width=\textwidth,
        trim={0 105 0 52},
        clip
    ]{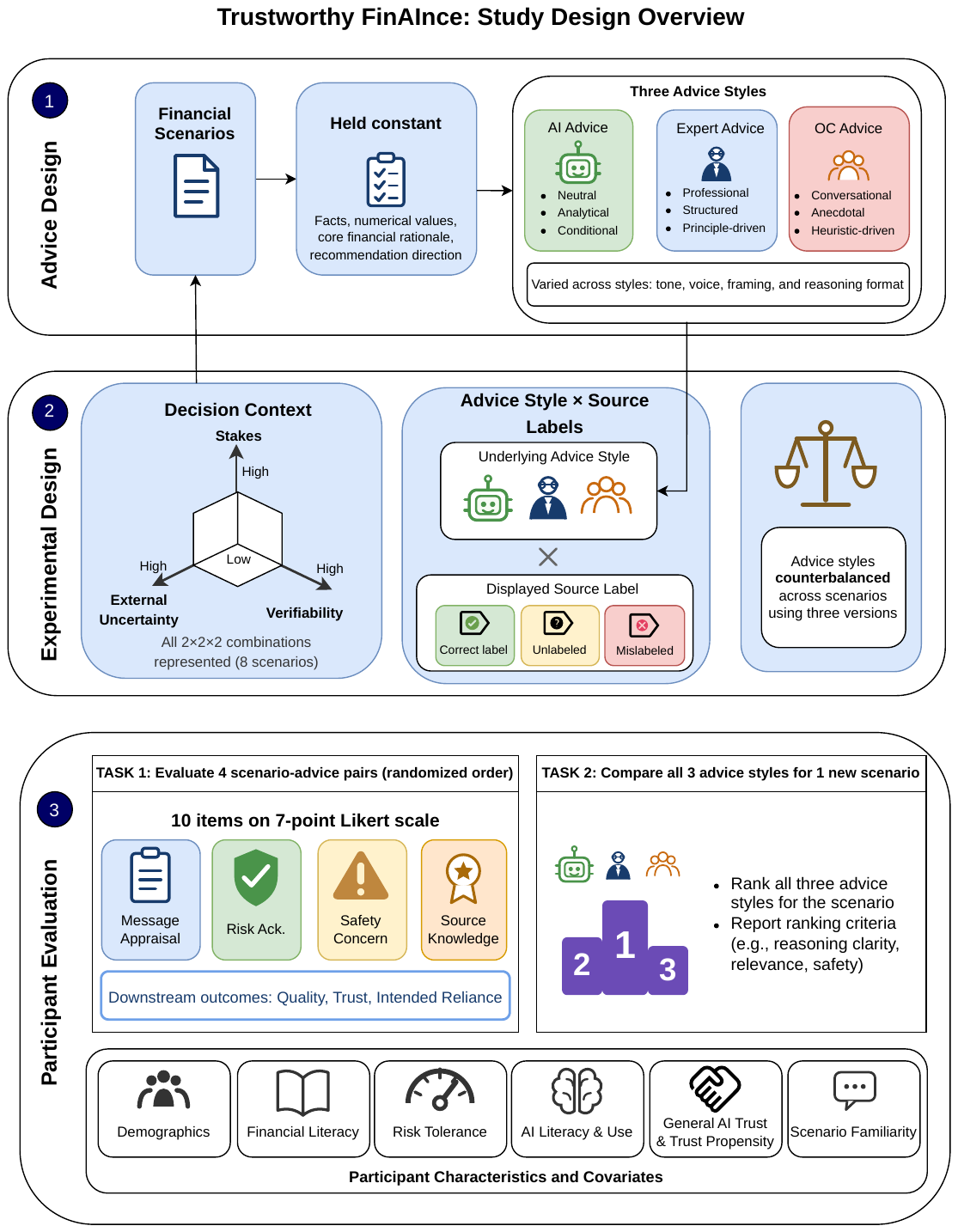}
    \Description{
Overview of the study design. Eight financial scenarios span all combinations
of stakes, external uncertainty, and verifiability. Advice is presented in AI,
Expert, and Online Community styles while scenario-specific facts, numerical
values, core financial rationale, and recommendation direction are held
constant. Source labels are independently manipulated as correctly labeled,
unlabeled, or mislabeled. Participants evaluate four scenario--advice pairs,
then rank all three advice styles for one previously unseen scenario. The study
also measures participant characteristics including financial literacy, risk
tolerance, AI literacy and use, trust, and scenario familiarity.
}
   \caption{
Overview of the study design. Eight financial scenarios represent all
combinations of stakes, external uncertainty, and verifiability. Advice was
constructed in AI, Expert, and Online Community (OC) styles while preserving
scenario-specific facts, numerical values, core financial rationale, and
recommendation direction. Source labels were independently manipulated as
\textit{correctly labeled}, \textit{unlabeled}, or \textit{mislabeled}.
Participants first evaluated four scenario--advice pairs and then ranked all
three advice styles for one previously unseen scenario. Advice styles were
counterbalanced across scenarios, and participant characteristics were also measured.
}
    \label{fig:study-design}
\end{figure}

We conducted a preregistered randomized vignette experiment to examine how
people evaluate personal financial advice as a function of \textit{advice
style} and \textit{source labels}~\cite{kapadia2026people}. Across eight financial scenarios, we
constructed advice in three styles associated with an AI Financial Assistant
(AI), Certified Financial Planner (Expert), and Online Community (OC).
AI-style advice was generated first and verified by the first author for
factual accuracy and alignment with the intended recommendation. Corresponding
Expert- and OC-style versions were then manually crafted while holding constant
the scenario-specific facts, numerical values, recommendation direction, and
core financial rationale, while varying source-associated communication style,
tone, and reasoning format. Source labels were independently manipulated as
matching (\textit{labeled condition}), absent (\textit{unlabeled condition}), or incorrect (\textit{mislabeled condition}). Participants first completed a repeated-measures evaluation task in which they rated four scenario--advice pairs on 10 items each. They subsequently completed a comparative ranking task in which all three advice styles were shown for a previously unseen scenario. Figure~\ref{fig:study-design} summarizes the complete experimental design.

\subsection{Participants, Recruitment, and Ethics}
\begin{table}[t]
\centering
\sffamily
\small

\caption{Participant demographics and background characteristics
($N=285$). Continuous variables are reported as mean (SD); categorical
variables are reported as $n$ (\%).}
\label{tab:participant_demographics_final}

\setlength{\tabcolsep}{7pt}
\renewcommand{\arraystretch}{1.13}

\begin{tabularx}{\textwidth}{
    >{\raggedright\arraybackslash}p{0.28\textwidth}
    >{\raggedright\arraybackslash}X
}
\toprule
\textbf{Characteristic} & \textbf{Final analytic sample} \\
\midrule

\multicolumn{2}{l}{\textit{\textbf{Demographics}}} \\[2pt]

Age
& 39.76 (12.22) \\

Sex
& Male: 143 (50.2\%); Female: 137 (48.1\%);
Prefer not to say: 3 (1.1\%); Unavailable: 2 (0.7\%) \\

Race/ethnicity
& White: 178 (62.5\%); Black: 41 (14.4\%);
Asian: 34 (11.9\%); Mixed: 15 (5.3\%);
Other: 10 (3.5\%); Prefer not to say: 3 (1.1\%);
Unavailable: 4 (1.4\%) \\

Household income
& Median category: \$50,000--\$74,999 \\

\addlinespace[4pt]
\multicolumn{2}{l}{\textit{\textbf{Financial Background}}} \\[2pt]

Objective financial literacy (0--3)
& 2.66 (0.68) \\

Self-rated financial literacy
& Very low: 2 (0.7\%); Low: 20 (7.0\%);
Moderate: 154 (54.0\%); High: 90 (31.6\%);
Very high: 19 (6.7\%) \\

Financial risk tolerance (1--4)
& No financial risk: 36 (12.6\%);
Average risk: 161 (56.5\%);
Above-average risk: 77 (27.0\%);
Substantial risk: 11 (3.9\%) \\

\addlinespace[4pt]
\multicolumn{2}{l}{\textit{\textbf{AI Experience and Attitudes}}} \\[2pt]

AI use frequency
& Never: 9 (3.2\%); Rarely: 20 (7.0\%);
Occasionally: 42 (14.7\%);
Regularly: 80 (28.1\%);
Frequently: 134 (47.0\%) \\

AI use for personal finance
& No, and have not considered it: 53 (18.6\%);
No, but have considered it: 43 (15.1\%);
Used once or twice: 102 (35.8\%);
Used multiple times: 87 (30.5\%) \\

General Trust in AI (1--7)
& 4.63 (1.50) \\

Human Trust Propensity (1--7)
& 4.61 (1.44) \\

Perceived AI Literacy (1--7)
& 5.16 (0.99) \\

\bottomrule
\end{tabularx}

\end{table}
Participants were recruited through Prolific. Eligibility was restricted to U.S. residents aged 18--70 who were fluent in English, had completed at least 100 previous Prolific submissions, and had an approval rate of at least 95\%.  The final analytic sample consisted of $N=285$ participants (\textit{labeled condition:} $n=100$, \textit{unlabeled condition:} $n=92$, and \textit{mislabeled condition:} $n=93$) after excluding incomplete responses, failed attention checks, and submissions completed in under two minutes, yielding 1,140 vignette-level observations in the primary evaluation task and 285 observations in the comparative ranking task. Table~\ref{tab:participant_demographics_final} summarizes participant
demographics and background characteristics. Participants received \$3 for completing the 15-minute survey. The study was reviewed and approved by the Institutional Review Board at our university.

\subsection{Financial Decision Scenarios}
\subsubsection{Decision-Context Taxonomy}
We constructed eight personal-finance scenarios using a $2\times2\times2$
factorial taxonomy spanning three complementary properties of financial
decisions: \textit{stakes}, \textit{external uncertainty}, and
\textit{verifiability}. \textit{Stakes} captures the magnitude and persistence
of a decision's financial consequences; high-stakes decisions involved larger
or longer-term consequences than low-stakes decisions.
\textit{External uncertainty} captures whether outcomes depend on unpredictable
future events, such as market movements or unexpected expenses; high-uncertainty
decisions depended more strongly on such events.
\textit{Verifiability} captures whether the quality of a recommendation can be
evaluated against an established financial principle; high-verifiability
decisions had a clearer normative benchmark than low-verifiability decisions which were preference-driven.
The eight scenarios represented all combinations of these three dimensions.
Experimental vignette methodology enables systematic variation of theoretically
relevant attributes while retaining concrete decision contexts
~\cite{aguinisBestPracticeRecommendations2014,
atzmullerExperimentalVignetteStudies2010}.

\subsubsection{Scenario Construction}
Each of the eight combinations of these dimensions was represented by a real-world financial scenario, covering decisions such as housing, credit card debt, travel insurance, emergency savings, graduate education, debt repayment and budgeting, long-term investing, and concentrated stock risk. Scenarios were approximately 100--150 words and described a protagonist's financial position, relevant numerical information, the competing considerations, and a concrete advice-seeking question. We operationalized the three dimensions before constructing the final advice stimuli and reviewed each scenario against the corresponding stakes, uncertainty, and verifiability criteria. Table~\ref{tab:scenario_taxonomy} in the appendix summarizes
the eight scenarios and their factorial assignments.

\subsection{Advice Stimulus Construction}

We first generated an AI-style advice response using a prompt that instructed the model (ChatGPT (OpenAI; GPT-5.3 Instant)) to act as a ``neutral AI financial assistant'' and respond to the financial decision described in the scenario. We then manually verified each response for factual accuracy and alignment with the
intended financial recommendation. Corresponding Expert- and Online Community (OC)-style responses were then manually crafted to preserve the scenario-specific facts, numerical
values, recommendation direction, and core financial rationale while varying
source-associated advice style, tone, and reasoning format. Construction was
informed by public financial-advice materials, including online financial
communities, financial-planning resources, and conversational AI responses.

The three conditions were designed to capture recurring communication patterns
associated with contemporary financial-advice sources rather than to represent
all communication from any source category. \textit{AI-style} advice used a neutral,
impersonal voice with explicit analytical and conditional reasoning.
Expert-style advice used a professional, measured tone with principle-based
reasoning, explicit trade-offs, and longer-term planning considerations.
OC-style advice used a conversational voice, experiential framing, informal
language, and practical heuristics. Table~\ref{tab:advice_style_operationalization}
summarizes the style specifications.

\subsubsection{Stimulus Validation}
\label{sec:stimulus_validation}

We conducted complementary linguistic, human, semantic, and external-reference
checks to assess whether the constructed responses reflected the intended
advice styles while preserving key substantive content. First, we compared 12
linguistic features associated with the intended style dimensions, including
lexical complexity, formality, personal voice, conditional language, hedging,
and contractions
~\cite{sahaAIVsHumans2026,tausczikPsychologicalMeaningWords2010,
coleman1975computer,babakovDontLoseMessage2023}. Using Friedman tests across
the eight matched scenario triplets with Benjamini--Hochberg FDR correction,
10 of 12 features differed across styles. The Categorical--Dynamic Index
(CDI)~\cite{pennebakerWhenSmallWords2014} showed a clear gradient from AI
($M=27.36$) to Expert ($M=21.30$) to OC ($M=12.03$),
$\chi^2(2)=14.25$, $p<.001$, with all pairwise contrasts remaining significant
after FDR correction.

Three independent annotators, blind to the intended style labels, then
classified each of the 24 advice texts and evaluated substantive consistency
within each scenario. The intended style was identified in 91.7\% of individual
judgments, with majority classification matching all 24 stimuli
(Fleiss' $\kappa=.749$). Across all eight scenario triplets, majority judgments
indicated preservation of the recommendation direction, scenario-specific
facts, and core financial rationale. Annotators flagged potentially material
additional information in five triplets, primarily in OC responses, typically
reflecting the anecdotal or experiential framing intentionally associated with
that style. As a complementary semantic check, Sentence-Transformers embeddings (\texttt{all-mpnet-base-v2})
~\cite{reimersSentenceBERTSentenceEmbeddings2019} showed substantially greater
similarity within than across scenarios (AI--Expert: $.844$ vs.\ $.304$; AI--OC: $.728$ vs.\ $.282$; Expert--OC: $.798$ vs.\ $.293$); across 24 directed cross-style
comparisons, the nearest semantic counterpart always came from the same
financial scenario (100\% top-1 retrieval).

\begin{figure*}[t]
    \centering
    \includegraphics[
        width=\textwidth
    ]{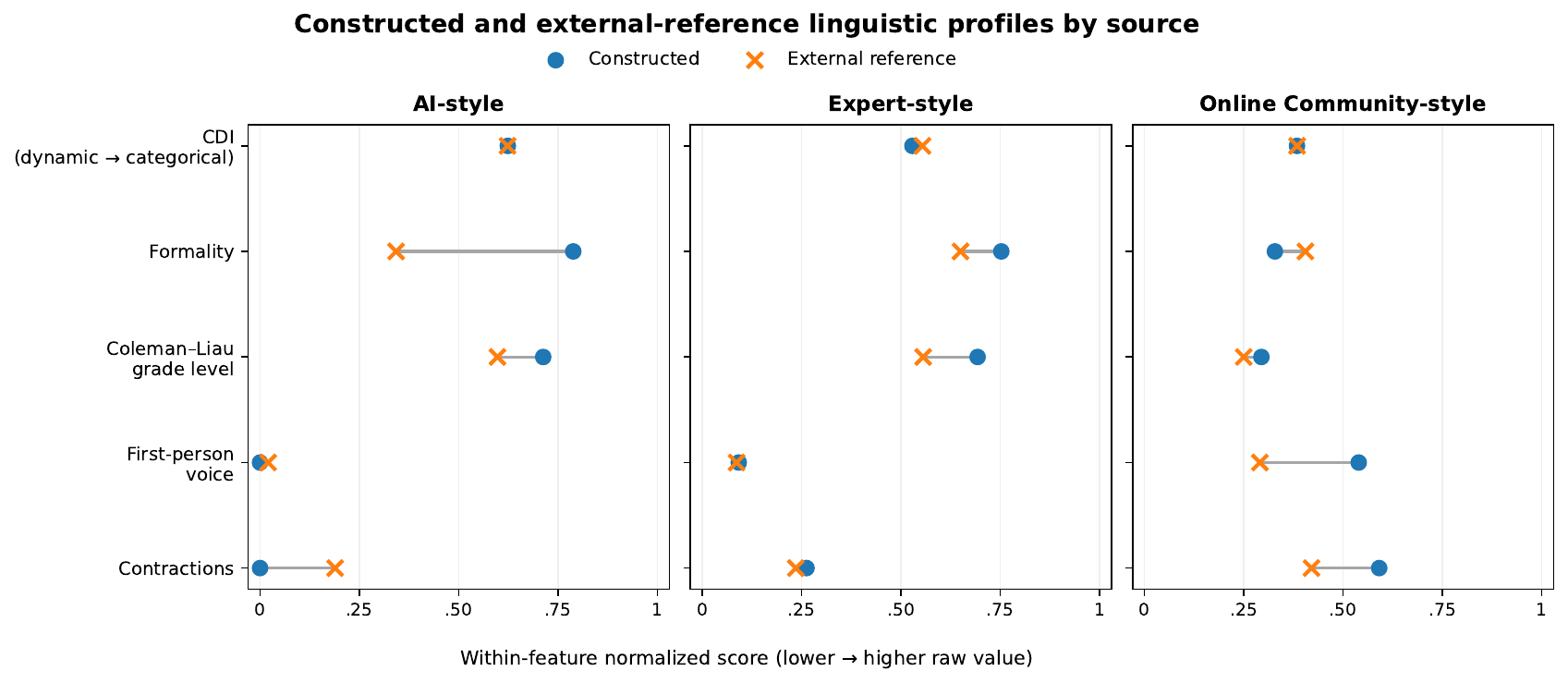}
    \Description{
    Correspondence between constructed and external-reference linguistic
    profiles across AI, Expert, and Online Community advice. Source-level
    means are normalized within each linguistic dimension using the combined
    distribution of constructed and reference texts. Circles denote constructed
    stimuli, crosses denote external-reference texts, and connecting-line length
    reflects the discrepancy between the two profiles within each feature.
    Higher values indicate more categorical language for CDI and higher raw
    values for the remaining measures. Normalized positions should be
    interpreted within, rather than across, linguistic dimensions.
    }
    \caption{
    Correspondence between constructed and external-reference linguistic
    profiles across AI, Expert, and Online Community advice. Source-level
    means are normalized within each linguistic dimension using the combined
    distribution of constructed and reference texts. Circles denote constructed
    stimuli, crosses denote external-reference texts, and connecting-line length
    reflects the discrepancy between the two profiles within each feature.
    Higher values indicate more categorical language for CDI and higher raw
    values for the remaining measures. Normalized positions should be
    interpreted within, rather than across, linguistic dimensions. Exact values
    and statistical comparisons are reported in
    Appendix~\ref{app:stimulus_validation_details}.
    }
    \label{fig:linguistic_profile_alignment}
\end{figure*}

Finally, we compared the constructed stimuli with a topic-matched reference
corpus of 72 naturally occurring financial-advice texts from AI systems,
financial professionals, and online communities. Several broad linguistic
patterns reproduced across the constructed and reference texts, including
differences in lexical complexity, first-person language, contractions, and
CDI, although individual cues such as formality, conditional language, and
hedging did not always follow the same ordering. We therefore interpret the
manipulation as capturing controlled \textit{source-associated advice styles},
rather than literal representations of all communication from each source
category. Figure~\ref{fig:linguistic_profile_alignment} summarizes the external
comparison.

\subsection{Experimental Manipulation and Assignment}

Participants were randomly assigned between subjects to one of three
source-label conditions. In the \textit{correctly labeled} condition, advice
was displayed with its corresponding source label (AI, Expert, or Online
Community); in the \textit{unlabeled} condition, no source label was shown; and
in the \textit{mislabeled} condition, advice was displayed with an incorrect
source label. Two counterbalanced mislabeling schemas were used. In Schema A, AI-, Expert-,
and OC-style advice were displayed with OC, AI, and Expert labels,
respectively, in Task~1, and with Expert, OC, and AI labels in Task~2.
Schema B reversed these mappings across the two tasks. When present, source
labels appeared before the advice.

The eight scenarios were divided into two groups of four. Participants were
randomly assigned to a scenario group and one of three counterbalanced
advice-style versions, such that each scenario appeared in each advice style
across participants. Each participant evaluated all three advice styles across
their four scenarios, with one style appearing twice, and scenario order was
randomized. For the comparative ranking task, participants received a randomly
selected, previously unseen scenario from the other scenario group and viewed
all three advice-style versions side by side.



\subsection{Procedure}

After providing informed consent, participants completed background questions
on financial advice seeking and participant characteristics. They then
completed four randomized scenario evaluations. For each, participants read a
financial vignette, viewed one advice response under their assigned source-label
condition, reported scenario familiarity, and rated the advice on ten evaluation
items. After an attention check, participants completed the comparative ranking
task using a previously unseen scenario from the other scenario group. They
viewed all three advice-style versions side by side, ranked them from most to
least preferred, and indicated the criteria that influenced their ranking.
The survey concluded with the remaining participant measures and debriefing.



\subsection{Measures}

\subsubsection{Advice Evaluation}

After each scenario--advice pair, participants rated ten items on seven-point
Likert scales assessing readability, reasoning clarity, situational fit, risk
acknowledgment, financial-harm risk, misleadingness, source knowledge, overall
quality, Trust, and intended Reliance. For analysis, readability, reasoning
clarity, and situational fit were combined as \textit{Message Appraisal};
financial-harm risk and misleadingness were combined as \textit{Safety
Concern}; and \textit{Risk Acknowledgment} and \textit{Source Knowledge} were
retained as single-item appraisals. \textit{Overall Quality}, \textit{Trust},
and  \textit{Intended Reliance} were retained as distinct downstream outcomes.
Trust captured whether participants would trust the advice in the
protagonist's situation, whereas intended Reliance captured whether they would
feel comfortable following it. Exact item wording appears in
Table~\ref{tab:measurement_items}.

\subsubsection{Participant Characteristics}

We measured participant characteristics that could shape financial-advice
evaluation. Objective financial literacy was assessed using the ``Big Three''
questions on compound interest, inflation, and diversification
~\cite{lusardiFinancialLiteracyWorld2011,
lusardiEconomicImportanceFinancial2014}, alongside self-rated financial
literacy. Financial risk tolerance was measured using the Survey of Consumer
Finances item~\cite{grableFinancialRiskTolerance1999}. AI-related measures
included AI-use frequency, prior AI use for personal finance, perceived AI
literacy (PAILQ-6)~\cite{grassiniPsychometricValidationPAILQ62024}, and
General Trust in AI~\cite{mcgrathMeasuringTrustArtificial2025}. We additionally
measured Human Trust Propensity
~\cite{frazierDevelopmentValidationPropensity2013}, prior use of financial
advice sources, and scenario-specific familiarity.


\subsubsection{Comparative Ranking Task} 
The comparative ranking task provided a direct within-scenario comparison of the three advice styles. Participants viewed the AI, Expert, and OC advice for one previously unseen scenario and ranked the three responses from most to least preferred. They then indicated the considerations that informed their ranking, including readability, clarity of reasoning, safety, situational relevance, overall reliability, and alignment with their own thinking or preferences. Participants also reported their confidence in their ranking on a 5-point scale.

\subsection{Analysis Strategy}

Because each participant evaluated four pieces of advice, our primary analyses
used linear mixed-effects models with participant-level random intercepts to
account for repeated observations. Advice-evaluation outcomes were standardized
before modeling. We first examined which upstream factors---advice style, source
labels, decision context, and scenario familiarity---were associated with the
four appraisals (RQ1). We then examined how these appraisals were associated
with \textit{Overall Quality}, \textit{Trust}, and \textit{Intended Reliance},
sequentially incorporating preceding evaluation variables to characterize the
broader evaluative pathway (RQ2). Table~\ref{tab:analysis_strategy} summarizes
the model specifications, inferential tests, and multiple-comparison
corrections used throughout.

We conducted two sets of complementary analyses. First, to disentangle advice
style from source labels, we tested whether their associations were additive or
interactive and decomposed their appraisal-carried pathways using
participant-level bootstrap resampling. Second, we examined potential boundary
conditions by testing both the direct associations of participant
characteristics and whether individual differences or decision context
moderated downstream evaluation pathways. Where multiple related hypotheses
were tested, we applied the correction procedures specified in
Table~\ref{tab:analysis_strategy}. Because appraisals and downstream judgments
were measured within the same advice-evaluation task, we interpret the pathway
analyses as associational rather than as evidence of causal mediation.

Finally, we analyzed the side-by-side ranking task separately as convergent
evidence using first-choice and complete-ranking comparisons. We also conducted
a series of robustness checks addressing within-person variation, model
specification, scenario parameterization, and composite construction. Additional statistical details and robustness results are provided in the appendix.

\begin{table*}[t]
\centering
\sffamily
\scriptsize

\caption{Summary of the analysis strategy.}
\label{tab:analysis_strategy}

\setlength{\tabcolsep}{4pt}
\renewcommand{\arraystretch}{1.18}

\begin{tabularx}{\textwidth}{
    >{\raggedright\arraybackslash}p{0.19\textwidth}
    >{\raggedright\arraybackslash}X
    >{\raggedright\arraybackslash}p{0.18\textwidth}
}
\toprule
\textbf{Analysis}
& \textbf{Model / Method}
& \textbf{Inference} \\
\midrule

\multicolumn{3}{l}{
\textit{\textbf{RQ1: What factors shape people's appraisals of financial advice?}}
} \\[2pt]

Appraisals
&
$\mathrm{Appraisal} \sim
\mathrm{Upstream\ Factors} +
(1|\mathrm{Participant})$
&
LMM; $\beta$, $p$
\\

Participant characteristics
&
$\mathrm{Appraisal} \sim
\mathrm{Upstream\ Factors} +
\mathrm{Participant\ Characteristics} +
(1|\mathrm{Participant})$
&
Joint model; BH-FDR
\\

\midrule

\multicolumn{3}{l}{
\textit{\textbf{RQ2: How are appraisals associated with Overall Quality, Trust, and intended Reliance?}}
} \\[2pt]

Upstream baseline
&
$\mathrm{Outcome} \sim
\mathrm{Upstream\ Factors} +
(1|\mathrm{Participant})$
&
$R^2_m$, $R^2_c$
\\

Overall Quality
&
$\mathrm{Overall\ Quality} \sim
\mathrm{Upstream\ Factors} +
\mathrm{Four\ Appraisals} +
(1|\mathrm{Participant})$
&
LMM; $\beta$, $p$, $R^2$
\\

Trust
&
$\mathrm{Trust} \sim
\mathrm{Upstream\ Factors} +
\mathrm{Four\ Appraisals} +
\mathrm{Overall\ Quality} +
(1|\mathrm{Participant})$
&
LMM; $\beta$, $p$, $R^2$
\\

Intended Reliance
&
$\mathrm{Reliance} \sim
\mathrm{Upstream\ Factors} +
\mathrm{Four\ Appraisals} +
\mathrm{Overall\ Quality} +
\mathrm{Trust} +
(1|\mathrm{Participant})$
&
LMM; $\beta$, $p$, $R^2$
\\

\midrule

\multicolumn{3}{l}{
\textit{\textbf{Disentangling Advice Style and Source Labels}}
} \\[2pt]

Style $\times$ Label
&
$\mathrm{Outcome} \sim
\mathrm{Base\ Model} +
\mathrm{Advice\ Style}\times\mathrm{Source\ Label}$
&
LRT; BH-FDR for contrasts
\\

Pathway decomposition
&
Advice Style / Source Label
$\rightarrow$ Appraisals
$\rightarrow$ Overall Quality
$\rightarrow$ Trust
$\rightarrow$ Reliance
&
2,000 bootstraps; 95\% CI
\\

\midrule

\multicolumn{3}{l}{
\textit{\textbf{Boundary Conditions}}
} \\[2pt]

Participant characteristics
&
$\mathrm{Trust} \sim
\mathrm{RQ2\ Trust\ Model} +
\mathrm{Participant\ Characteristics}$

\smallskip
$\mathrm{Reliance} \sim
\mathrm{RQ2\ Reliance\ Model} +
\mathrm{Participant\ Characteristics}$
&
$\beta$, $p$; BH-FDR
\\

\addlinespace[2pt]

Individual-difference moderation
&
$\mathrm{Trust} \sim
\mathrm{Upstream\ Factors} +
\mathrm{Evaluation\ Predictors} +
\mathrm{Moderator} +
\mathrm{Moderator}\times\mathrm{Evaluation\ Predictors} +
(1|\mathrm{Participant})$

\smallskip
$\mathrm{Reliance} \sim
\mathrm{Upstream\ Factors} +
\mathrm{Evaluation\ Predictors} +
\mathrm{Trust} +
\mathrm{Moderator} +
\mathrm{Trust}\times\mathrm{Moderator} +
(1|\mathrm{Participant})$
&
LRT $\chi^2$; BH-FDR
\\

\addlinespace[2pt]

Decision-context moderation
&
$\mathrm{Trust} \sim
\mathrm{Upstream\ Factors} +
\mathrm{Evaluation\ Predictors} +
\mathrm{Context}\times\mathrm{Evaluation\ Predictors} +
(1|\mathrm{Participant})$

\smallskip
$\mathrm{Reliance} \sim
\mathrm{Upstream\ Factors} +
\mathrm{Evaluation\ Predictors} +
\mathrm{Trust} +
\mathrm{Context}\times
(\mathrm{Evaluation\ Predictors}+\mathrm{Trust}) +
(1|\mathrm{Participant})$
&
LRT $\chi^2$; BH-FDR
\\

\midrule

\multicolumn{3}{l}{
\textit{\textbf{Other Analyses}}
} \\[2pt]

Comparative ranking
&
First-choice frequencies and complete within-participant rankings
&
$\chi^2$; Friedman; Kendall's $W$; Wilcoxon + Holm
\\

Robustness
&
Within--between decomposition; GEE; scenario fixed effects;
constituent-item models
&
Sensitivity checks
\\

\bottomrule
\end{tabularx}

\vspace{3pt}
\begin{minipage}{0.96\textwidth}
\footnotesize
\textit{Note.}
\textit{Upstream Factors} comprise Advice Style, Source Label, Stakes,
External Uncertainty, Verifiability, and Scenario Familiarity.
\textit{Four Appraisals} comprise Message Appraisal, Risk Acknowledgment,
Safety Concern, and Source Knowledge.
\textit{Evaluation Predictors} comprise the Four Appraisals and Overall Quality.
\textit{Participant Characteristics} comprise the seven individual-difference
measures entered jointly.
\textit{Moderator} denotes one focal participant characteristic, and
\textit{Context} denotes one focal decision-context dimension
(Stakes, External Uncertainty, or Verifiability).
\textit{Base Model} denotes the corresponding additive model before the focal
interaction.
All mixed-effects models include participant-level random intercepts.
BH-FDR denotes Benjamini--Hochberg false-discovery-rate correction;
LRT denotes likelihood-ratio test.
\end{minipage}

\end{table*}


\section{Results}

\subsection{Organizing Advice-Evaluation Measures into Appraisals}

We organized the 10 advice-evaluation items according to the role they play in the advice-evaluation process. Building on the Information Adoption Model (IAM), which distinguishes evaluations of the message from evaluations of its source~\cite{sussmanInformationalInfluenceOrganizations2003}, we averaged \textit{readability}, \textit{reasoning clarity}, and \textit{situational fit} into a \textit{Message Appraisal} index ($\alpha=.887$). These items capture how understandable, well-reasoned, and applicable the advice appears. We retained \textit{Source Knowledge} separately as a source-credibility judgment. Because financial advice can appear clear and credible while still presenting potential downside or epistemic risk, we evaluated risk and safety separately, consistent with prior financial-AI work that distinguishes information quality from perceived risk in advice evaluation~\cite{ioannouRoleExplainabilityAIDriven2026}. \textit{Risk of financial harm} and \textit{misleadingness} were averaged into a \textit{Safety Concern} index ($\alpha=.712$; inter-item $r=.55$), with higher values indicating greater concern. \textit{Risk Acknowledgment} was retained separately because recognizing potential risks in the advice is conceptually different from judging the advice itself as harmful or misleading. Table~\ref{tab:composite_reliability} presents information on the four appraisals. Finally, we retained \textit{Overall Quality}, \textit{Trust}, and \textit{Intended Reliance} as distinct downstream judgments, allowing us to examine how these initial appraisals relate to broader evaluation, trust, and willingness to follow the advice.

\begin{table}[t]
\centering
\sffamily
\footnotesize

\caption{Construction and internal consistency of the four advice appraisals across the 1,140 advice evaluations. Higher \textit{Safety Concern} scores indicate greater perceived risk of financial harm or misleadingness. Dashes indicate statistics that are not applicable to single-item appraisals.}
\label{tab:composite_reliability}

\begin{tabular}{p{0.23\linewidth} p{0.40\linewidth} c c c}
\toprule
\textbf{Appraisal}
& \textbf{Items}
& $\boldsymbol{k}$
& $\boldsymbol{\alpha}$
& \textbf{Mean inter-item $r$} \\
\midrule

Message Appraisal
& Readability; reasoning clarity; situational fit
& 3
& .887
& .72 \\

Risk Acknowledgment
& Risk acknowledgment
& 1
& --
& -- \\

Safety Concern
& Financial-harm risk; misleadingness
& 2
& .712
& .55 \\

Source Knowledge
& Perceived source knowledge
& 1
& --
& -- \\

\bottomrule
\end{tabular}
\end{table}

\subsection{Advice Style, Source Labels, and Decision Context Shape Different Appraisals}

Advice style, source labels, and decision context were associated with different appraisal profiles (\autoref{tab:appraisal-models}).

\paragraph{Advice style primarily shaped Message Appraisal and Safety Concern.}

Compared with AI-style advice, both Expert- and Online Community (OC)-style advice received higher \textit{Message Appraisal} ratings ($\beta_{\text{Expert}}=.334$, $p<.001$; $\beta_{\text{OC}}=.223$, $p<.001$) and lower \textit{Safety Concern} ($\beta_{\text{Expert}}=-.245$, $p<.001$; $\beta_{\text{OC}}=-.127$, $p=.026$). The styles also differed in \textit{Source Knowledge}: Expert-style advice was perceived as coming from a more knowledgeable source than AI-style advice ($\beta=.141$, $p=.025$), whereas OC-style advice was perceived as less knowledgeable ($\beta=-.174$, $p=.006$), producing an Expert--AI--OC ordering on perceived source knowledge. \textit{Risk acknowledgment} showed a different pattern: Expert and AI styles did not significantly differ ($\beta=-.061$, $p=.340$), while OC-style advice was rated lower than AI-style advice ($\beta=-.147$, $p=.021$)

\paragraph{Source labels produced more selective differences.}

Relative to unlabeled advice, an Expert label was associated with greater \textit{Source Knowledge} ($\beta=.225$, $p=.017$) and higher \textit{Message Appraisal} ($\beta=.190$, $p=.035$). AI and OC labels did not differ significantly from the unlabeled condition across the four appraisals. Thus, source labels were most clearly reflected in perceived source expertise.

\paragraph{Decision context primarily shaped risk and safety evaluations.}

High-stakes scenarios increased \textit{Safety Concern} ($\beta=.436$, $p<.001$) and reduced \textit{Message Appraisal} ($\beta=-.241$, $p<.001$). Greater external uncertainty increased \textit{Risk Acknowledgment} ($\beta=.310$, $p<.001$), while higher verifiability reduced both \textit{Risk Acknowledgment} ($\beta=-.199$, $p=.010$) and \textit{Safety Concern} ($\beta=-.292$, $p<.001$). Scenario familiarity was associated with more favorable evaluations across several dimensions: greater Message Appraisal ($\beta=.319$, $p<.001$), greater Risk Acknowledgment ($\beta=.140$, $p<.001$), greater perceived Source Knowledge ($\beta=.206$, $p<.001$), and lower Safety Concern ($\beta=-.125$, $p<.001$).

\paragraph{Individual differences shaped selected appraisals.}
Using the same base model, we examined participant characteristics jointly and found selective associations with the four appraisals.(Table~\ref{tab:individual-appraisal-main}). Greater AI Literacy was associated with higher \textit{Message Appraisal} ($\beta=.189$, $p_{\mathrm{FDR}}<.001$) and \textit{Source Knowledge} ($\beta=.178$, $p_{\mathrm{FDR}}<.001$). General AI Trust ($\beta=.168$, $p_{\mathrm{FDR}}=.001$) and Objective Financial Literacy ($\beta=.086$, $p_{\mathrm{FDR}}=.038$) were also positively associated with \textit{Source Knowledge}. Greater Risk Tolerance was associated with higher \textit{Safety Concern} ($\beta=.145$, $p_{\mathrm{FDR}}=.015$) and lower \textit{Source Knowledge} ($\beta=-.117$, $p_{\mathrm{FDR}}=.011$). Overall, individual differences affected particular appraisals rather than shifting advice appraisal uniformly.

Overall, these results reveal differentiated entry points into advice appraisal: advice style was most broadly and consistently associated with message and safety appraisals, Expert source labels with perceived source knowledge, decision context with risk and safety appraisals, and individual differences with selected appraisal dimensions.

\begin{table*}[t]
\centering
\sffamily
\footnotesize
\caption{Mixed-effects models predicting the four initial appraisals of financial advice.
Coefficients are standardized fixed-effect estimates ($\beta$). Models include a participant-level random intercept. AI-style advice and the unlabeled condition are the reference categories. Higher \textit{Safety Concern} indicates greater perceived potential for financial harm or misleadingness.}
\label{tab:appraisal-models}

\setlength{\tabcolsep}{8pt}

\begin{tabular}{lcccc}
\toprule
&
\makecell{\textbf{Message}\\\textbf{Appraisal}} &
\makecell{\textbf{Risk}\\\textbf{Ack.}} &
\makecell{\textbf{Safety}\\\textbf{Concern$^\dagger$}} &
\makecell{\textbf{Source}\\\textbf{Knowledge}} \\
\midrule

\multicolumn{5}{l}{\textit{Advice style}} \\
\addlinespace[0.15em]

Expert vs.\ AI
& \posthree{+0.334***}
& -0.061
& \negthree{-0.245***}
& \posone{+0.141*} \\

OC vs.\ AI
& \posthree{+0.223***}
& \negone{-0.147*}
& \negone{-0.127*}
& \negtwo{-0.174**} \\

\addlinespace[0.45em]
\multicolumn{5}{l}{\textit{Source labels}} \\
\addlinespace[0.15em]

AI label vs.\ Unlabeled
& +0.075
& -0.088
& -0.141
& +0.049 \\

Expert label vs.\ Unlabeled
& \posone{+0.190*}
& -0.026
& -0.166
& \posone{+0.225*} \\

OC label vs.\ Unlabeled
& +0.168
& -0.044
& -0.119
& +0.091 \\

\addlinespace[0.45em]
\multicolumn{5}{l}{\textit{Decision context}} \\
\addlinespace[0.15em]

High stakes
& \negthree{-0.241***}
& +0.083
& \posthree{+0.436***}
& +0.010 \\

High external uncertainty
& -0.072
& \posthree{+0.310***}
& +0.026
& -0.005 \\

High verifiability
& +0.099
& \negtwo{-0.199**}
& \negthree{-0.292***}
& +0.036 \\

\addlinespace[0.45em]
\multicolumn{5}{l}{\textit{Covariate}} \\
\addlinespace[0.15em]

Scenario familiarity
& \posthree{+0.319***}
& \posthree{+0.140***}
& \negthree{-0.125***}
& \posthree{+0.206***} \\

\addlinespace[0.25em]
\midrule

Marginal $R^2$ ($R^2_m$)
& 0.170 & 0.054 & 0.115 & 0.067 \\

Conditional $R^2$ ($R^2_c$)
& 0.405 & 0.298 & 0.458 & 0.312 \\

\bottomrule
\end{tabular}

\vspace{0.5em}

\begin{minipage}{0.95\textwidth}
\footnotesize
\textit{Note.} $N=1{,}140$ evaluations from 285 participants.
Scenario familiarity is standardized; binary experimental predictors are coded 0/1.
$^{*}p<.05$, $^{**}p<.01$, $^{***}p<.001$.
\end{minipage}

\end{table*}
\begin{table}[t]
\centering
\sffamily
\footnotesize
\caption{Participant characteristics associated with initial advice appraisals.}
\label{tab:individual-appraisal-main}

\setlength{\tabcolsep}{6pt}
\renewcommand{\arraystretch}{1.15}

\begin{tabular}{
    p{0.35\linewidth}
    p{0.29\linewidth}
    rr
}
\textbf{Participant characteristic}
& \textbf{Appraisal}
& $\boldsymbol{\beta}$
& $\boldsymbol{p_{\mathrm{FDR}}}$ \\
\toprule

AI Literacy
& Message Appraisal
& .189 & $<$.001 \\

AI Literacy
& Source Knowledge
& .178 & $<$.001 \\

General AI Trust
& Source Knowledge
& .168 & .001 \\

Objective Financial Literacy
& Source Knowledge
& .086 & .038 \\

Risk Tolerance
& Safety Concern
& .145 & .015 \\

Risk Tolerance
& Source Knowledge
& $-.117$ & .011 \\

\bottomrule
\end{tabular}

\vspace{0.35em}
\begin{minipage}{0.96\linewidth}
\footnotesize
\textit{Note.}
Each appraisal was modeled separately using a linear mixed-effects model
with participant-level random intercepts. Models included the same base predictors
(advice style, source labels, decision context, and scenario familiarity),
with all seven participant characteristics entered simultaneously.
Benjamini--Hochberg FDR correction was applied across the seven
participant-characteristic coefficients within each appraisal.
Only associations surviving $p_{\mathrm{FDR}}<.05$ are shown.
\end{minipage}
\end{table}

\subsection{From Appraisals to Quality, Trust, and Intended Reliance}

We next examined how the four appraisals were associated with \textit{Overall Quality}, \textit{Trust}, and intended \textit{Reliance}. For each outcome, we compared an upstream model containing advice style, source labels, decision context, and scenario familiarity with a model that additionally included the preceding appraisal variables (\autoref{tab:upstream-evaluation-models}).

\paragraph{Appraisals were strongly associated with Overall Quality.}
All four intermediate appraisals were independently associated with \textit{Overall Quality}. \textit{Source Knowledge} showed the strongest association ($\beta=.489$, $p<.001$), followed by \textit{Message Appraisal} ($\beta=.325$, $p<.001$). Greater \textit{Safety Concern} predicted lower \textit{Overall Quality} ($\beta=-.088$, $p<.001$), while greater \textit{Risk Acknowledgment} predicted somewhat higher quality ratings ($\beta=.075$, $p<.001$). Adding the appraisals also increased marginal explained variance from 9.8\% to 69.2\%. We also observe that initial Expert-style and Expert-label associations with Overall Quality were substantially attenuated after accounting for these appraisals.

\paragraph{Overall Quality was the strongest correlate of Trust.}
Once the appraisals and \textit{Overall Quality} were included, \textit{Overall Quality} was by far the strongest correlate of \textit{Trust} ($\beta=.607$, $p<.001$). \textit{Message Appraisal} ($\beta=.132$, $p<.001$), lower \textit{Safety Concern} ($\beta=-.125$, $p<.001$), and \textit{Source Knowledge} ($\beta=.109$, $p<.001$) retained smaller independent associations with Trust, whereas \textit{Risk Acknowledgment} did not. Marginal explained variance increased from 11.2\% in the upstream model to 75.9\% in the appraisal- and quality-adjusted model. The initial Expert-style association with Trust was also no longer detectable after including these variables.

\paragraph{Trust was the strongest correlate of Intended Reliance.}
\textit{Trust} was the dominant correlate of  \textit{Intended Reliance} ($\beta=.594$, $p<.001$), followed by \textit{Overall Quality} ($\beta=.206$, $p<.001$); the remaining appraisals showed substantially smaller associations: \textit{Message Appraisal} ($\beta=.079$, $p<.001$), lower \textit{Safety Concern} ($\beta=-.050$, $p<.001$), and \textit{Source Knowledge} ($\beta=.047$, $p=.020$). Adding the full evaluation pathway increased marginal explained variance from 14.8\% to 82.9\%. After accounting for these variables, initial advice-style and source-label differences in Reliance were substantially reduced. That said, because we measured these intermediate judgments within the same advice evaluation, we interpret this structure as an \textit{evaluative pathway} rather than evidence of causal mediation.


Across all three downstream outcomes, models including the appraisal pathway explained substantially more variance than upstream factors alone, while the initial advantages associated with Expert advice style and Expert source labels were largely attenuated after accounting for these appraisals. Taken together, the results are consistent with a progressively organized evaluative pathway in which advice style, source labels, and decision context first shape distinct message, risk, safety, and source-related appraisals; these appraisals are strongly associated with \textit{Overall Quality}; \textit{Overall Quality} is in turn strongly associated with \textit{Trust}; and \textit{Trust} is the strongest correlate of \textit{Intended Reliance}.

\begin{table*}[t]
\centering
\sffamily
\footnotesize
\caption{Upstream and pathway-adjusted models of Overall Quality,
Trust, and Intended Reliance.}
\label{tab:upstream-evaluation-models}

\setlength{\tabcolsep}{5.5pt}
\renewcommand{\arraystretch}{0.8}

\begin{adjustbox}{width=\textwidth}
\begin{tabular}{lcccccc}
\toprule
&
\multicolumn{2}{c}{\textbf{Overall Quality}}
&
\multicolumn{2}{c}{\textbf{Trust}}
&
\multicolumn{2}{c}{\textbf{Intended Reliance}} \\

\cmidrule(lr){2-3}
\cmidrule(lr){4-5}
\cmidrule(lr){6-7}

\textbf{Predictor}
& \textbf{Upstream}
& \makecell{\textbf{Pathway-}\\\textbf{adjusted}}
& \textbf{Upstream}
& \makecell{\textbf{Pathway-}\\\textbf{adjusted}}
& \textbf{Upstream}
& \makecell{\textbf{Pathway-}\\\textbf{adjusted}} \\
\midrule

\multicolumn{7}{l}{\textit{Advice style}} \\
\addlinespace[0.15em]

Expert vs.\ AI
& \posthree{+0.205***}
& +0.011
& \postwo{+0.185**}
& -0.028
& \postwo{+0.173**}
& -0.019 \\

OC vs.\ AI
& -0.004
& +0.005
& +0.009
& -0.009
& +0.023
& +0.002 \\

\addlinespace[0.45em]
\multicolumn{7}{l}{\textit{Source labels}} \\
\addlinespace[0.15em]

AI label vs.\ Unlabeled
& +0.126
& +0.074
& +0.066
& -0.038
& +0.103
& +0.029 \\

Expert label vs.\ Unlabeled
& \postwo{+0.248**}
& +0.064
& +0.176
& -0.046
& \posone{+0.210*}
& +0.013 \\

OC label vs.\ Unlabeled
& +0.170
& +0.060
& +0.154
& +0.004
& \posone{+0.189*}
& +0.040 \\

\addlinespace[0.45em]
\multicolumn{7}{l}{\textit{Decision context}} \\
\addlinespace[0.15em]

High stakes
& -0.040
& \posone{+0.069*}
& \negone{-0.107*}
& -0.000
& \negtwo{-0.155**}
& -0.048 \\

High external uncertainty
& -0.013
& -0.008
& +0.001
& +0.014
& +0.017
& +0.030 \\

High verifiability
& +0.152
& \posone{+0.091*}
& \posone{+0.166*}
& +0.026
& \posone{+0.178*}
& +0.022 \\

\addlinespace[0.45em]
\multicolumn{7}{l}{\textit{Covariate}} \\
\addlinespace[0.15em]

Scenario familiarity
& \posthree{+0.256***}
& \posone{+0.038*}
& \posthree{+0.280***}
& \posone{+0.040*}
& \posthree{+0.324***}
& \posthree{+0.049***} \\

\addlinespace[0.25em]
\midrule
\multicolumn{7}{l}{\textit{Appraisal-to-reliance pathway}} \\
\addlinespace[0.15em]

Message Appraisal
& ---
& \posthree{+0.325***}
& ---
& \posthree{+0.132***}
& ---
& \posthree{+0.079***} \\

Risk Acknowledgment
& ---
& \posthree{+0.075***}
& ---
& +0.024
& ---
& -0.015 \\

Safety Concern$^\dagger$
& ---
& \negthree{-0.088***}
& ---
& \negthree{-0.125***}
& ---
& \negthree{-0.050***} \\

Source Knowledge
& ---
& \posthree{+0.489***}
& ---
& \posthree{+0.109***}
& ---
& \posone{+0.047*} \\

Overall Quality
& ---
& ---
& ---
& \posthree{+0.607***}
& ---
& \posthree{+0.206***} \\

Trust
& ---
& ---
& ---
& ---
& ---
& \posthree{+0.594***} \\

\addlinespace[0.25em]
\midrule
\multicolumn{7}{l}{\textit{Model fit}} \\
\addlinespace[0.10em]

Marginal $R^2$ ($R^2_m$)
& 0.098
& \textbf{0.692}
& 0.112
& \textbf{0.759}
& 0.148
& \textbf{0.829} \\

Conditional $R^2$ ($R^2_c$)
& 0.373
& 0.742
& 0.351
& 0.795
& 0.362
& 0.829 \\

\bottomrule
\end{tabular}
\end{adjustbox}

\vspace{0.5em}

\begin{minipage}{0.98\textwidth}
\footnotesize
\textit{Note.}
Cells report standardized fixed-effect coefficients ($\beta$) from
linear mixed-effects models with participant-level random intercepts.
Upstream models include advice style, source labels, decision context,
and scenario familiarity. Pathway-adjusted models additionally include
the four initial appraisals; the Trust model further includes Overall Quality,
and the Reliance model further includes Overall Quality and Trust.
AI-style advice and the unlabeled condition are the reference categories.
Models include $N=285$ participants and $1{,}140$ advice evaluations.
$R^2_m$ and $R^2_c$ denote marginal and conditional $R^2$, respectively.
Orange and blue shading indicate significant positive and negative coefficients,
respectively; shading intensity reflects statistical significance
($p<.05$, $p<.01$, and $p<.001$), not effect magnitude.
Unshaded coefficients are not statistically significant.
$^\dagger$ Higher values indicate greater perceived financial harm or
misleadingness.
$^{*}p<.05$, $^{**}p<.01$, $^{***}p<.001$.
\end{minipage}

\end{table*}

\subsection{Advice Style and Source Labels Operate Additively Through Different Appraisals}
\label{sec:style-label-pathways}

In a subsequent exploratory analysis, we tested whether the associations of advice style depended on the source label displayed with the advice. Across the four appraisals and three downstream outcomes, adding Advice Style $\times$ Source Label interactions did not improve model fit (all $p>.29$; Table~\ref{tab:style-label-interactions}). Thus, these results support treating advice style and displayed source labels as largely additive rather than interactive effects.

We then examined whether these two cues were reflected differently through the appraisal pathway. Participant-level bootstrap decompositions showed that advice style was relatively more strongly reflected through \textit{Message Appraisal}, whereas source labels were relatively more
strongly reflected through \textit{Source Knowledge}. For Expert advice, the Message-versus-Knowledge pathway was more pronounced for advice style than for the corresponding Expert label
($\Delta\Delta=.081$, 95\% CI $[.014,.148]$); the same pattern appeared for Online Community advice ($\Delta\Delta=.124$, 95\% CI $[.045,.212]$). These results suggest that how advice is written is reflected more strongly in appraisals of the message itself, whereas explicit source labels shift appraisal relatively toward perceived source expertise. Full pathway decompositions and simple contrasts are reported in Table~\ref{tab:pathway_decomposition} and Table~\ref{tab:pathway_contrasts}, respectively.

\subsection{Boundary Conditions of the Evaluation Pathway}
\label{sec:pathway-boundaries}

We next examined whether the relationships among appraisals, \textit{Overall Quality}, \textit{Trust}, and  \textit{Intended Reliance} varied across decision contexts or participant characteristics.

\paragraph{Decision context did not substantially alter the downstream pathway.}
Although stakes, external uncertainty, and verifiability shaped the initial appraisals participants formed, we found no robust evidence that any of the three decision-context dimensions moderated how those appraisals translated into \textit{Trust} or intended \textit{Reliance} after FDR correction. Thus, decision context appeared to matter primarily at the appraisal stage rather than in the subsequent evaluation pathway.

\paragraph{Individual differences produced selective moderation.}

Objective Financial Literacy strengthened the associations of \textit{Source Knowledge} with \textit{Trust} ($\beta_{\mathrm{int}}=.054$, $p_{\mathrm{FDR}}=.017$) and of \textit{Trust} with \textit{Intended Reliance} ($\beta_{\mathrm{int}}=.039$, $p_{\mathrm{FDR}}=.002$). Human Trust Propensity selectively moderated the association between \textit{Safety Concern} and \textit{Trust} ($\beta_{\mathrm{int}}=.056$, $p_{\mathrm{FDR}}=.002$), with safety concerns weighing more strongly among participants lower in trust propensity. General AI Trust was associated with higher overall\textit{ Trust} ($\beta=.202$, $p_{\mathrm{FDR}}<.001$) and \textit{Intended Reliance} ($\beta=.227$, $p_{\mathrm{FDR}}<.001$), but these associations were substantially reduced after accounting for the appraisal pathway.

Overall, the downstream pathway was comparatively stable across decision contexts, while individual differences altered only selected parts of how appraisals translated into Trust and Reliance. More details are reported in Appendix~\ref{sec:appendix-moderation}.

\subsection{Comparative Ranking Provides Convergent Evidence}
\label{sec:ranking}

\begin{figure}[t]
    \centering
    \includegraphics[width=0.8\linewidth]{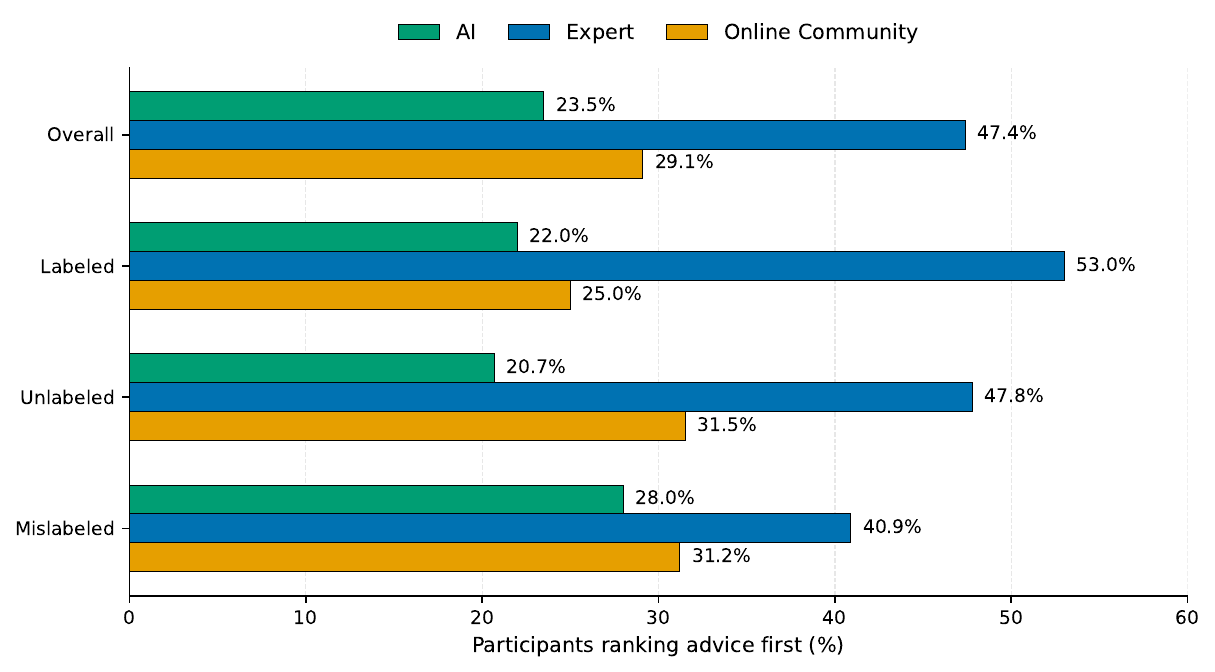}
    \Description{
    Percentage of participants ranking each advice style first in the
comparative ranking task, overall and by source label condition.
Expert-style advice was most frequently ranked first across all conditions.
Although its descriptive advantage was smaller in the Mislabeled condition,
the overall Advice Style $\times$ Attribution Arm interaction was not
statistically significant.
    }
    \caption{Percentage of participants ranking each advice style first in the
comparative ranking task, overall and by source label condition.
Expert-style advice was most frequently ranked first across all conditions.
Although its descriptive advantage was smaller in the Mislabeled condition,
the overall Advice Style $\times$ Source Label Arm interaction was not
statistically significant.}
    \label{fig:ranking_percent}
\end{figure}

As a final comparative task, participants ranked AI-, Expert-, and Online
Community (OC)-style advice side by side for a previously unseen financial
scenario. Expert-style advice was ranked first by 47.4\% of participants,
compared with 29.1\% for OC advice and 23.5\% for AI advice
($\chi^2(2)=26.61$, $p<.001$). Across the complete rankings, the three advice styles also differed significantly (Friedman $\chi^2(2)=37.16$, $p<.001$,
Kendall's $W=.065$). Pairwise Wilcoxon signed-rank tests with Holm
correction showed that Expert-style advice was preferred over both
AI- and OC-style advice (both $p_{\mathrm{Holm}}<.001$), whereas
AI and OC did not differ.
Notably, the Expert-style preference remained in the Unlabeled condition,
where 47.8\% ranked it first, compared with 31.5\% for OC and 20.7\% for AI (Figure~\ref{fig:ranking_percent}).
We found no evidence that the relative ranking of the three styles
systematically differed across source-label conditions
($\chi^2(4)=6.65$, $p=.155$).

\begin{figure}[t]
    \centering
    \includegraphics[width=0.8\linewidth]{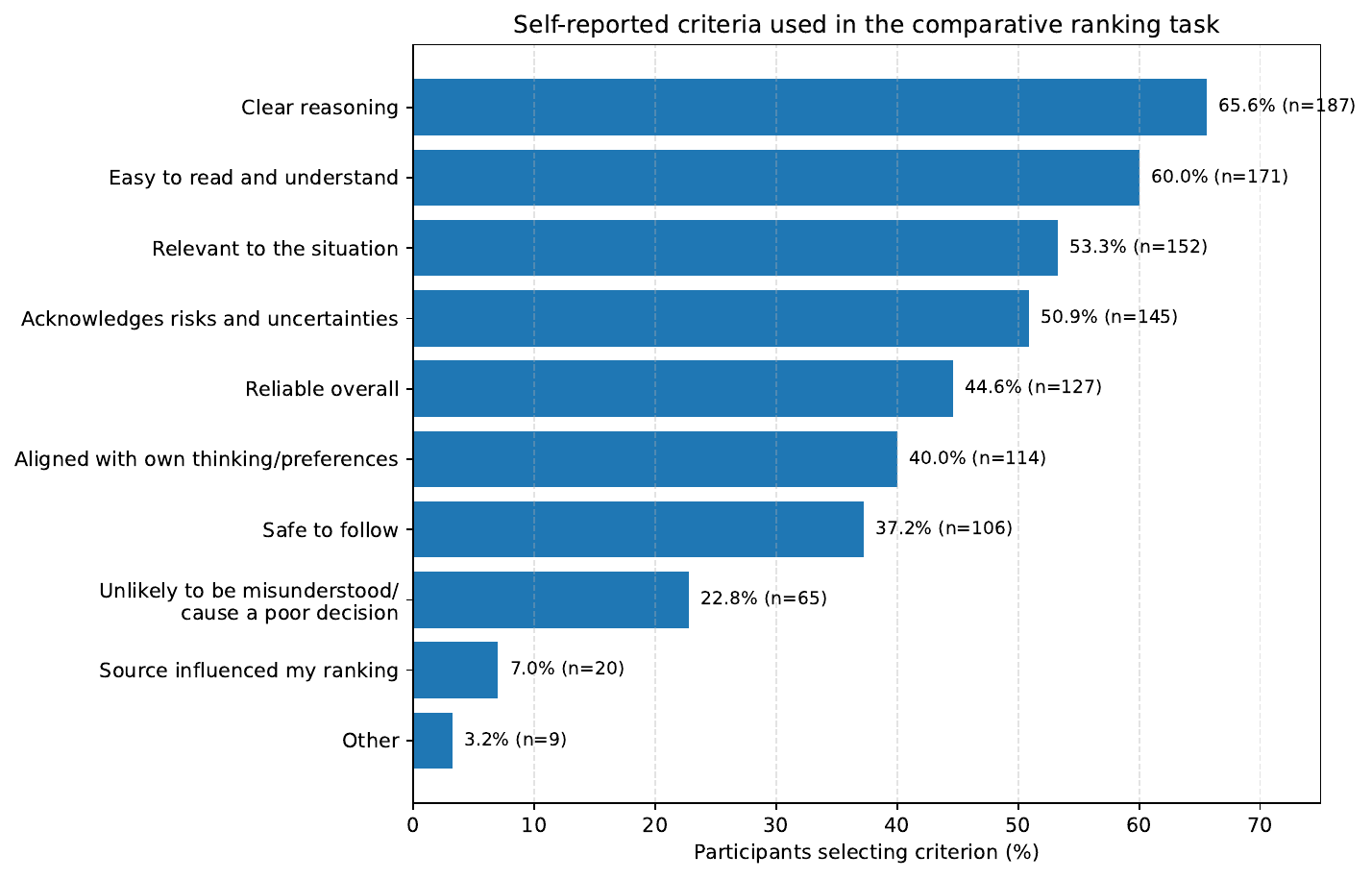}
    \Description{Self-reported criteria used in the comparative ranking task. Participants could select multiple criteria. Percentages indicate the proportion of participants who selected each criterion. Clear reasoning, readability, situational relevance, and acknowledgment of risks and uncertainties were the most frequently reported considerations, whereas explicit source influence was comparatively rare.}
    \caption{Self-reported criteria used in the comparative ranking task. Participants could select multiple criteria. Percentages indicate the proportion of participants who selected each criterion. Clear reasoning, readability, situational relevance, and acknowledgment of risks and uncertainties were the most frequently reported considerations, whereas explicit source influence was comparatively rare.}
    \label{fig:ranking_criteria}
\end{figure}

Participants' self-reported ranking criteria further emphasized properties
of the advice itself. Clear reasoning (65.6\%), ease of reading and
understanding (60.0\%), situational relevance (53.3\%), and acknowledgment
of risks and uncertainties (50.9\%) were the most frequently selected
criteria, whereas only 7.0\% reported that the displayed source influenced
their ranking (\autoref{fig:ranking_criteria}). Although these self-reports
do not identify what causally determined participants' rankings, the
comparative task provides convergent evidence for the primary results:
participants preferred Expert-style advice and predominantly described
their preferences in terms of message-level appraisal cues rather than
explicit source information.

\subsection{Robustness Checks}

We conducted four robustness checks to assess whether the appraisal pathway depended on stable participant response tendencies, the repeated-measures estimator, the representation of decision context, or the construction of the appraisal composites. First, within--between participant decompositions showed that the principal associations remained at the within-participant level, indicating that the pathway was not driven solely by participants who generally rated all advice more favorably. Second, generalized estimating equations closely reproduced the mixed-effects estimates; for example, the \textit{Overall Quality}--\textit{Trust} association was $\beta=.608$ (vs.\ $.607$ in the primary model) and the \textit{Trust}--\textit{Reliance} association was $\beta=.595$ (vs.\ $.594$). Third, replacing the decision-context dimensions with scenario fixed effects produced nearly identical pathway estimates. Finally, replacing \textit{Message Appraisal} and \textit{Safety Concern} with their constituent items preserved the main conclusions. Overall, the appraisal--Quality--Trust--Reliance pattern was stable across alternative specifications and measurement choices. More details are reported in Appendix~\ref{sec:appendix-robustness}. 
\section{Discussion \label{section:discussion}}

The landscape of personal financial decision-making is rapidly evolving as generative AI becomes an increasingly accessible source of financial guidance. As these systems become more prevalent, the central challenge is not simply whether people trust AI-generated advice, but whether that trust and subsequent reliance are grounded in an appropriate evaluation of the advice. 
Existing trust-calibration approaches commonly rely on cues such as disclaimers, source labels, communication choices, and anthropomorphic design, but examining their direct relationship with Trust provides only a partial account of how users arrive at these judgments. 
Our work instead frames trust and reliance as part of a broader evaluation process, showing that advice style, source labels, and decision context are associated with distinct appraisals of the message, its risks and safety, and the source's perceived knowledge. 
We discuss the implications of this work.

\subsection{Extending Information Adoption for Financial Advice Evaluation}

A growing body of work on AI-generated information and human--AI advisory
systems has approached users' responses through the lens of information
evaluation and adoption~\cite{ioannouRoleExplainabilityAIDriven2026,
liHowUsersAdopt2025,gongWhenAlgorithmsSpeak2026,
camilleriAcceptanceUsageChatGPT2024}. Classical IAM accounts, as discussed
previously, organize this process around evaluations of \textit{argument
quality} and \textit{source credibility}, which shape perceived usefulness
and subsequent adoption~\cite{sussmanInformationalInfluenceOrganizations2003}.
Our work extends this perspective by opening up the process through which
financial advice becomes trusted and relied upon. 
Rather than treating
\textit{Trust} as a direct response to an advice source or interface cue, our
results characterize it as part of a broader evaluation process in which
people form distinct appraisals of the message, its risks and safety, and its
source, which are associated with \textit{Overall Quality}, \textit{Trust},
and ultimately \textit{Intended Reliance}. 
This more granular view matters
because similar levels of trust can mask very different underlying appraisal
profiles. For example, Online-Community-style advice was appraised more
favorably than AI-style advice on message and safety dimensions, but less
favorably on perceived \textit{Source Knowledge}; these differences largely
offset downstream, leaving trust in the two styles statistically
indistinguishable. Looking beneath aggregate trust therefore provides a more
diagnostic account of \emph{why} advice is trusted and which message-, risk-,
safety-, or source-related cues may warrant design attention, consistent with
work on appropriate reliance that does not treat higher Trust as inherently
desirable~\cite{leeTrustAutomationDesigning2004,raeesPeopleAppropriatelyRely2026}.

Our findings also elaborate this evaluation process for financial decisions
by making risk- and safety-related appraisals explicit. Financial advice can
appear clear, relevant, and credible while still failing to acknowledge
uncertainty or exposing users to potential harm. Recent work on AI-assisted
financial decision-making similarly highlights perceived risk and uncertainty
as important to advice adoption~\cite{ioannouRoleExplainabilityAIDriven2026}.
Our results show that these appraisals were sensitive to characteristics of
the decision itself: greater external uncertainty was associated with greater
\textit{Risk Acknowledgment} ($\beta=.310$), while higher stakes were
associated with greater \textit{Safety Concern} ($\beta=.436$); higher
verifiability was associated with lower levels of both. These patterns suggest that perceptions of risk and safety are closely associated with the characteristics of the financial decision in which the advice is encountered. Decision context therefore appears to enter the
evaluation process by influencing which aspects of uncertainty and potential
harm become salient to users, extending information-evaluation accounts beyond
properties of the message and its source alone.

Finally, our findings reinforce the distinction between \textit{Trust} and
\textit{Intended Reliance}. Trust is commonly treated as an antecedent of
reliance on automated systems, but the two are not equivalent:
\citet{leeTrustAutomationDesigning2004} emphasize that the central design
problem is whether trust supports \emph{appropriate reliance}, while
advice-taking research similarly distinguishes attitudes toward a source from
whether its recommendation is incorporated into a decision
~\cite{bonaccioAdviceTakingDecisionmaking2006}. In our results,
\textit{Trust} was the strongest correlate of \textit{Intended Reliance}, yet
\textit{Overall Quality}, \textit{Message Appraisal}, and
\textit{Safety Concern} retained smaller independent associations after
accounting for Trust. Participant-level variation further supported this
distinction: scenario familiarity was directly associated with greater
Reliance ($\beta=.089$, $p<.001$) despite a weaker association with Trust,
while objective financial literacy strengthened the Trust--Reliance
relationship ($\beta_{\mathrm{int}}=.039$, $p_{\mathrm{FDR}}=.002$).
Thus, similar levels of \textit{Trust} need not translate into similar
intentions to act. This aligns with prior work suggesting that studies of
AI-supported decision-making should measure Trust and Reliance separately and,
where possible, complement stated intentions with behavioral measures of
whether users accept, reject, or verify advice
~\cite{raeesPeopleAppropriatelyRely2026}.

\subsection{Advice Style and Source Labels as distinct epistemic cues}

Prior HCI research shows that how AI communicates can shape how
users evaluate and trust it. Authoritative language can increase Trust even
when system limitations are disclosed~\cite{metzgerEmpoweringCalibratedDistrust2024},
while social and personality cues such as friendliness, extroversion, and
stance adaptation can similarly influence responses to AI
~\cite{takayanagiAreGenerativeAI2025,sunBeFriendlyNot2026}. Parallel work
shows that displayed source information can act as a separate credibility
cue: identical information may be evaluated differently depending on whether
it is attributed to AI, a human, or an expert
~\cite{jakeschAIMediatedCommunicationHow2019,reisInfluenceBelievedAI2024,
sunUnderstandingTrustHuman2026,sahebiAIPenaltyDisclosure2026}. Our findings bring these two strands together by independently varying advice style and source labels, showing that they were associated with distinct appraisal profiles. Advice style was reflected more broadly in message-
and safety-related appraisals. Consistent with this pattern, Expert-style advice remained the most frequently preferred in the comparative ranking task
even when no source label was displayed. Expert source labels, in contrast,
were most clearly associated with perceived \textit{Source Knowledge}. Their
associations were also largely additive rather than interactive, suggesting
that sounding like an expert and being identified as an expert provide
distinct rather than interchangeable signals.

From an AI-governance perspective, this distinction raises concerns around
provenance and accountability. Generative AI can readily reproduce the
linguistic markers of professional advice---such as structured reasoning,
measured language, and explicit trade-offs---without necessarily carrying the
expertise, review, or responsibility associated with a qualified professional.
This gap between \emph{presentational authority} and \emph{epistemic
authority} may encourage unwarranted confidence when polished communication is
mistaken for evidence of expertise. Our findings therefore suggest that advice
style and source information should be treated as distinct layers of evaluation and governance
in AI-assisted financial advice. Systems should make clear who produced,
reviewed, or endorsed a recommendation and where responsibility lies, rather
than allowing professional presentation alone to signal authority. This distinction is already relevant in deployed systems: \textit{SoFi} reports that its financial planners were involved in evaluating responses from Coach, illustrating how AI-generated advice may incorporate human professional review without being directly authored by a human advisor~\cite{WhatSoFiLearned}.

\subsection{Design Implications for AI-Mediated Financial Advice}

Our findings suggest three considerations for the design and evaluation of
AI-mediated financial advice systems.

\paragraph{Design for grounded evaluation and appropriate Reliance.}
Financial AI systems should help users assess \emph{why} a recommendation
deserves Trust and whether it warrants action, rather than optimizing Trust
through surface-level cues such as professionalized advice style alone. This challenge is already visible in deployed financial AI systems: in testing
its AI financial guide, Coach, \textit{SoFi} reported that effective financial guidance
depends not only on the underlying financial logic, but also on how recommendations
and trade-offs are communicated to users~\cite{WhatSoFiLearned}. Interfaces could make relevant assumptions, reasoning, trade-offs, uncertainty, and potential downside consequences easier to inspect---for example, through expandable assumptions or a brief ``Why this
recommendation?'' view. Such support should ultimately help users decide when
to follow, verify, reject, or escalate a recommendation, including by facilitating comparison with other sources or suggesting professional consultation when appropriate. Evaluations of these systems should likewise distinguish \textit{Trust} from \textit{Intended Reliance} and, where possible, examine actual advice-taking and verification behavior rather than positive perceptions alone.

\paragraph{Separate professional presentation from verifiable epistemic authority.}
Clear and professional communication can improve how advice is evaluated, but
it should not substitute for evidence of expertise. Financial advice systems
should distinguish whether advice was \textit{AI-generated},
\textit{professionally reviewed}, or \textit{directly authored by a qualified
advisor}; make the relevant expertise or credentials behind that involvement
visible; expose the evidence or financial principles supporting the
recommendation; and clarify where responsibility lies and what limitations
apply. Provenance could function as an independently inspectable layer
alongside the advice itself, rather than as a decorative ``Expert'' badge.
This separation can preserve the benefits of clear communication without
allowing presentational authority alone to stand in for warranted epistemic
authority.

\paragraph{Adapt evaluation support to the decision and the user.}

 Decision context was reflected primarily in the appraisals users formed; interfaces could therefore foreground different information depending on what makes a financial decision difficult to evaluate. For example,high-stakes decisions might emphasize potential downside consequences, uncertain decisions might make assumptions and contingencies explicit, and difficult-to-verify decisions could make trade-offs and alternative courses of action more explicit rather than presenting a single recommendation as definitive. User-sensitive support could similarly help people evaluate cues such as source expertise or safety concerns, rather than assuming that all users interpret these cues in the same way. Lastly, because greater familiarity was consistently associated with more favorable appraisals of trust and Reliance, financial AI systems may need to guard against familiarity reducing scrutiny. Even when a scenario feels familiar to the user, interfaces should preserve lightweight checks on assumptions, risks, and verifiability so reliance is based on evaluating the advice rather than familiarity with the situation alone.

\subsection{Limitations and Future Work}

Our work has several limitations that also suggest interesting future
directions. First, \textit{Reliance} was measured as behavioral intention rather
than observed financial behavior. Real financial decisions often unfold over
time, involve monetary consequences, and include consultation with other people
or information sources. Future work should therefore examine whether the
evaluation patterns observed here predict actual advice taking, verification,
information seeking, and decision outcomes, including whether users
appropriately accept, reject, or escalate recommendations.

Second, although our results are consistent with an evaluative organization
linking appraisals, \textit{Overall Quality}, \textit{Trust}, and
\textit{Intended Reliance}, these constructs were measured within the same advice-
evaluation task. The observed associations therefore should not be interpreted
as establishing a causal or temporal pathway. Future work could test this
ordering more directly through longitudinal or experimental designs that
manipulate specific appraisals and observe subsequent changes in Trust and
Reliance.

Third, we used controlled financial scenarios and deliberately constructed
advice styles. This allowed us to hold the underlying recommendation relatively
constant while varying its presentation and source label, but no small set of
styles can capture the diversity of AI, professional, or online-community
advice. Real-world AI advice is also interactive: users ask follow-up questions,
introduce new constraints, challenge recommendations, and observe how systems
respond. Extending this work to multi-turn and longitudinal interactions is
therefore an important next step.

Fourth, our decision-context dimensions were instantiated through eight specific
financial scenarios. Although robustness analyses suggest that the downstream
pattern was not driven by individual scenarios, other financial decisions may
involve factors not represented here, including urgency, irreversibility,
regulatory complexity, social consequences, or ambiguity in user goals.
Similarly, our sample consisted of U.S. adults recruited through Prolific;
financial norms, access to professional advice, and expectations of AI and
expertise may differ across populations and institutional settings. Replication
across financial domains, countries, and real advisory settings will therefore
be important for establishing generalizability.

Finally, individual characteristics such as financial literacy, AI literacy,
and risk tolerance were measured rather than experimentally manipulated.
Accordingly, their associations and moderation patterns should not be
interpreted causally. Future work could test whether interventions that improve
financial knowledge, source-verification skills, or risk comprehension change
how users appraise advice and translate Trust into Reliance over time.
Longitudinal studies could also examine how these relationships evolve as users
accumulate experience with an AI system and observe the consequences of its
advice. 

\section{Conclusion}

As generative AI becomes an increasingly common source of personal finance advice, the challenge is not simply whether people trust its advice, but whether that trust is grounded in how they evaluate it. In a randomized experiment with 285 U.S. adults, we show that advice style, source labels, and decision context shaped distinct appraisals of the
message, its risks and safety, and its source. These appraisals were strongly associated with \textit{Overall Quality}, which was in turn closely associated with \textit{Trust}, while Trust was the strongest correlate of \textit{Intended Reliance}. Advice style and source labels contributed different signals: style was associated more broadly with message- and safety-related
appraisals, whereas Expert labels were most clearly associated with perceived \textit{Source Knowledge}. Decision context and individual differences further shaped selected parts of this evaluation process. Together, these findings shift the design goal from making financial AI appear trustworthy toward supporting grounded evaluation: helping users assess the reasoning, risks,
source expertise, and context of advice so that Trust and Reliance better reflect what the recommendation warrants.





\bibliographystyle{ACM-Reference-Format}
\bibliography{references_zotero,referencesCleaned}

%
\clearpage
\appendix
\providecommand{\FloatBarrier}{\clearpage}

\setcounter{table}{0}
\renewcommand{\thetable}{A\arabic{table}}

\section{Study Design Details}
\label{app:stimulus-details}

\begin{table}[h]
\centering
\sffamily
\small

\caption{Decision-context assignments for the eight financial scenarios.}
\label{tab:scenario_taxonomy}

\setlength{\tabcolsep}{5pt}
\renewcommand{\arraystretch}{1.12}

\begin{tabular}{lccc}
\toprule
\textbf{Scenario}
& \textbf{Stakes}
& \textbf{External Uncertainty}
& \textbf{Verifiability} \\
\midrule

Renting: Shared vs.\ Private
& Low & Low & Low \\

Credit Card vs.\ New Laptop
& Low & Low & High \\

Travel Insurance
& Low & High & Low \\

Emergency Fund vs.\ Investing Bonus
& Low & High & High \\

Graduate School vs.\ Continuing Work
& High & Low & Low \\

High-Interest Debt Repayment and Budgeting
& High & Low & High \\

Long-Term Investment Planning
& High & High & Low \\

Concentrated Stock Position Risk
& High & High & High \\

\bottomrule
\end{tabular}
\end{table}

\begin{table}[h]
\centering
\sffamily
\small
\caption{Operationalization of the three source-associated advice styles.}
\label{tab:advice_style_operationalization}
\setlength{\tabcolsep}{5pt}
\renewcommand{\arraystretch}{1.15}
\begin{tabularx}{\textwidth}{
    >{\raggedright\arraybackslash}p{0.12\textwidth}
    >{\raggedright\arraybackslash}X
    >{\raggedright\arraybackslash}X
    >{\raggedright\arraybackslash}X
}
\toprule
\textbf{Feature} & \textbf{AI advice} & \textbf{Expert advice} & \textbf{OC advice} \\
\midrule
\textbf{Voice}
& Impersonal; no first-person identity
& Mild first-person professional authority
& First-person peer perspective \\

\textbf{Tone}
& Neutral and analytical
& Professional and measured
& Informal and conversational \\

\textbf{Reasoning}
& Explicit analytical and conditional reasoning
& Principle-based reasoning with explicit trade-offs and longer-term planning
& Experiential reasoning supported by practical heuristics \\

\textbf{Framing}
& Balances available options and financial risk exposure
& Connects the decision to financial foundations and broader goals
& Uses lived experience, subjective judgment, and relatable consequences \\

\textbf{Structure}
& Situation and trade-off; analysis; conditional recommendation
& Quantified framing; trade-off explanation; actionable principle
& Position or anecdote; experiential lesson; direct recommendation \\
\bottomrule
\end{tabularx}

\vspace{2pt}
\begin{minipage}{0.97\textwidth}
\footnotesize
\textit{Note.} These profiles were designed as controlled, source-associated
styles rather than literal representations of all communication from each
source category. Across versions of a scenario, scenario-specific facts,
numerical values, recommendation direction, and core financial rationale were
preserved while style, tone, and reasoning presentation varied.
\end{minipage}
\end{table}

\label{app:stimulus_validation_details}

\begin{table}[h]
\centering
\sffamily
\footnotesize
\caption{Selected linguistic profiles of the constructed stimuli and the
external reference corpus. Values are means.}
\label{tab:validation_profiles}
\setlength{\tabcolsep}{5pt}
\renewcommand{\arraystretch}{1.12}
\begin{tabular}{lrrrrrr}
\toprule
& \multicolumn{2}{c}{\textbf{AI}}
& \multicolumn{2}{c}{\textbf{Expert}}
& \multicolumn{2}{c}{\textbf{Online Community}} \\
\cmidrule(lr){2-3}\cmidrule(lr){4-5}\cmidrule(lr){6-7}
\textbf{Measure}
& \textbf{Constructed} & \textbf{Reference}
& \textbf{Constructed} & \textbf{Reference}
& \textbf{Constructed} & \textbf{Reference} \\
\midrule
Categorical--Dynamic Index & 27.36 & 27.31 & 21.30 & 22.92 & 12.03 & 12.05 \\
Formality & .824 & .530 & .801 & .733 & .521 & .572 \\
Coleman--Liau grade level & 13.67 & 12.04 & 13.39 & 11.45 & 7.74 & 7.11 \\
First-person singular / 100 words & 0.00 & 0.14 & 0.62 & 0.59 & 3.66 & 1.97 \\
Contractions / 100 words & 0.00 & 1.15 & 1.59 & 1.43 & 3.58 & 2.55 \\
\bottomrule
\end{tabular}

\vspace{2pt}
\begin{minipage}{0.97\textwidth}
\footnotesize
\textit{Note.} Constructed values summarize eight stimuli per advice style;
reference values summarize 24 topic-matched texts per source category. Higher
CDI values indicate more categorical/analytical language, whereas lower values
indicate more dynamic/interpersonal language. The main text reports the
inferential validation tests and the dimensions that did not reproduce the same
ordering across constructed and reference texts.
\end{minipage}
\end{table}

\begin{table}[h]
\centering
\sffamily
\small

\caption{Advice-evaluation items and corresponding analytic constructs.}
\label{tab:measurement_items}

\renewcommand{\arraystretch}{1.18}
\setlength{\tabcolsep}{5pt}

\begin{tabularx}{\textwidth}{
    >{\centering\arraybackslash}p{0.04\textwidth}
    >{\raggedright\arraybackslash}X
    >{\raggedright\arraybackslash}p{0.18\textwidth}
    >{\raggedright\arraybackslash}p{0.19\textwidth}
}
\toprule
\textbf{No.}
& \textbf{Measurement Item (7-point Likert scale)}
& \textbf{Item Construct}
& \textbf{Analytic Construct} \\
\midrule

\multicolumn{4}{l}{\textit{\textbf{Appraisals}}} \\[2pt]

1
& The advice is easy to read and follow.
& Readability
& Message Appraisal \\

2
& The advice provides clear reasoning for its recommendation.
& Reasoning Clarity
& Message Appraisal \\

3
& The advice fits well with X's situation.
& Situational Fit
& Message Appraisal \\

\addlinespace[2pt]

4
& The advice acknowledges potential risks and uncertainties.
& Risk Acknowledgment
& Risk Acknowledgment \\

\addlinespace[2pt]

5
& Following this advice could put X at financial risk.
& Financial Harm Risk
& Safety Concern$\dagger$ \\

6
& Someone with limited financial knowledge could misunderstand this advice.
& Misleadingness
& Safety Concern$\dagger$ \\

\addlinespace[2pt]

7
& The source of this advice appears knowledgeable about financial decisions.
& Source Knowledge
& Source Knowledge \\

\addlinespace[5pt]

\multicolumn{4}{l}{\textit{\textbf{Downstream Judgments}}} \\[2pt]

8
& Overall, this is high-quality financial advice for X.
& Overall Quality
& Overall Quality \\

9
& I would trust this advice if I were in X's situation.
& Trust
& Trust \\

10
& If I were in X's situation, I would feel comfortable following this advice.
& Intended Reliance
& Intended Reliance \\

\bottomrule
\end{tabularx}

\vspace{3pt}
\begin{minipage}{\textwidth}
\footnotesize
\textit{Note.} X denotes the protagonist named in each financial scenario.
Readability, Reasoning Clarity, and Situational Fit were combined to form
\textit{Message Appraisal}; Financial Harm Risk and Misleadingness were
combined to form \textit{Safety Concern}. \textit{Risk Acknowledgment} and
\textit{Source Knowledge} were retained as single-item appraisals.
Items marked with a dagger ($\dagger$) are negatively valenced, such that
higher ratings indicate greater safety concern.
\end{minipage}

\end{table}

\FloatBarrier

\section{Supplementary Statistical Results}
\label{app:statistical-results}

\subsection{Advice Style and Source Labels}

\begin{table}[h]
\sffamily
\centering
\small
\caption{Likelihood-ratio tests comparing additive mixed-effects models with
models including the Advice Style $\times$ Source Label interaction.}
\label{tab:style-label-interactions}
\setlength{\tabcolsep}{8pt}
\renewcommand{\arraystretch}{1.12}
\begin{tabular}{lrr}
\toprule
\textbf{Outcome} & $\boldsymbol{\chi^2(6)}$ & $\boldsymbol{p}$ \\
\midrule
Message Appraisal       & 3.71 & .716 \\
Risk Acknowledgment     & 7.32 & .293 \\
Safety Concern          & 2.20 & .901 \\
Source Knowledge        & 3.29 & .772 \\
Overall Quality         & 4.83 & .566 \\
Trust                   & 1.48 & .961 \\
Reliance                & 2.17 & .904 \\
\bottomrule
\end{tabular}
\end{table}

\FloatBarrier

\begin{table}[h]
\centering
\sffamily
\small
\caption{Bootstrap decomposition of appraisal-carried pathways from advice
style and source labels to intended Reliance. Bolded estimates have 95\%
bootstrap confidence intervals excluding zero.}
\label{tab:pathway_decomposition}
\setlength{\tabcolsep}{3.5pt}
\renewcommand{\arraystretch}{1.16}
\begin{tabular}{lccccc}
\toprule
\textbf{Upstream contrast}
& \makecell{\textbf{Message}\\\textbf{Appraisal}}
& \makecell{\textbf{Risk}\\\textbf{Ack.}}
& \makecell{\textbf{Safety}\\\textbf{Concern}}
& \makecell{\textbf{Source}\\\textbf{Knowledge}}
& \textbf{Total} \\
\midrule
\multicolumn{6}{l}{\textit{Advice style}} \\
Expert vs.\ AI
& \makecell{\textbf{.114}\\\textbf{[.069,.166]}}
& \makecell{-.003\\{}[-.009,.002]}
& \makecell{\textbf{.043}\\\textbf{[.023,.068]}}
& \makecell{\textbf{.055}\\\textbf{[.009,.106]}}
& \makecell{\textbf{.210}\\\textbf{[.115,.309]}} \\[0.35em]
OC vs.\ AI
& \makecell{\textbf{.076}\\\textbf{[.033,.123]}}
& \makecell{-.006\\{}[-.017,.001]}
& \makecell{\textbf{.022}\\\textbf{[.003,.045]}}
& \makecell{\textbf{-.068}\\\textbf{[-.129,-.014]}}
& \makecell{.023\\{}[-.083,.129]} \\
\midrule
\multicolumn{6}{l}{\textit{Source labels}} \\
AI vs.\ Unlabeled
& \makecell{.025\\{}[-.033,.090]}
& \makecell{-.003\\{}[-.014,.004]}
& \makecell{.024\\{}[-.009,.061]}
& \makecell{.018\\{}[-.053,.095]}
& \makecell{.064\\{}[-.072,.209]} \\[0.35em]
Expert vs.\ Unlabeled
& \makecell{\textbf{.064}\\\textbf{[.001,.129]}}
& \makecell{-.001\\{}[-.011,.008]}
& \makecell{.028\\{}[-.006,.066]}
& \makecell{\textbf{.088}\\\textbf{[.014,.164]}}
& \makecell{\textbf{.179}\\\textbf{[.032,.325]}} \\[0.35em]
OC vs.\ Unlabeled
& \makecell{\textbf{.058}\\\textbf{[.001,.118]}}
& \makecell{-.002\\{}[-.011,.006]}
& \makecell{.020\\{}[-.015,.056]}
& \makecell{.038\\{}[-.036,.114]}
& \makecell{.114\\{}[-.022,.256]} \\
\bottomrule
\end{tabular}
\vspace{2pt}
\begin{minipage}{0.97\textwidth}
\footnotesize
\textit{Note.} Cells report bootstrap mean pathway estimates with 95\%
percentile confidence intervals from 2,000 participant-level bootstrap
samples. AI is the advice-style reference; Unlabeled is the source-label
reference. These estimates are descriptive pathway decompositions rather than
causal mediation effects.
\end{minipage}
\end{table}

\FloatBarrier

\begin{table}[h]
\centering
\sffamily
\footnotesize
\caption{Bootstrap contrasts comparing the Message Appraisal and Source
Knowledge pathways to intended Reliance.}
\label{tab:pathway_contrasts}
\setlength{\tabcolsep}{6pt}
\renewcommand{\arraystretch}{1.15}
\begin{tabular}{p{0.45\columnwidth}cc}
\toprule
\textbf{Pathway contrast} & \textbf{Estimate} & \textbf{95\% CI} \\
\midrule
\multicolumn{3}{l}{\textit{Expert}} \\
Expert style: Message Appraisal $-$ Source Knowledge
& \textbf{.060} & \textbf{[.006,.112]} \\
Expert label: Message Appraisal $-$ Source Knowledge
& -.021 & [-.081,.035] \\
Style--label contrast-of-contrasts
& \textbf{.081} & \textbf{[.014,.148]} \\
\midrule
\multicolumn{3}{l}{\textit{Online Community}} \\
OC style: Message Appraisal $-$ Source Knowledge
& \textbf{.144} & \textbf{[.101,.190]} \\
OC label: Message Appraisal $-$ Source Knowledge
& .020 & [-.043,.082] \\
Style--label contrast-of-contrasts
& \textbf{.124} & \textbf{[.045,.212]} \\
\bottomrule
\end{tabular}
\vspace{2pt}
\begin{minipage}{0.95\columnwidth}
\footnotesize
\textit{Note.} Positive values indicate a relatively stronger Message
Appraisal pathway than Source Knowledge pathway. Bolded intervals exclude zero.
\end{minipage}
\end{table}

\FloatBarrier

\subsection{Boundary Conditions}
\label{sec:appendix-moderation}

\begin{table}[h]
\centering
\sffamily
\small
\caption{Supplementary tests of boundary conditions for Trust and intended
Reliance.}
\label{tab:boundary-condition-tests}
\setlength{\tabcolsep}{6pt}
\renewcommand{\arraystretch}{1.14}
\begin{tabular}{p{0.27\textwidth}p{0.35\textwidth}ccc}
\toprule
\textbf{Boundary condition} & \textbf{Tested association}
& \textbf{Statistic} & \textbf{df} & $\boldsymbol{p_{\mathrm{FDR}}}$ \\
\midrule
\multicolumn{5}{l}{\textit{Decision-context moderation of Trust}} \\
Stakes & Context $\times$ evaluation-pathway block & $\chi^2=11.52$ & 5 & .104 \\
External Uncertainty & Context $\times$ evaluation-pathway block & $\chi^2=10.23$ & 5 & .104 \\
Verifiability & Context $\times$ evaluation-pathway block & $\chi^2=5.94$ & 5 & .312 \\
\addlinespace[0.3em]
\multicolumn{5}{l}{\textit{Decision-context moderation of intended Reliance}} \\
Stakes & Context $\times$ evaluation-pathway block & $\chi^2=5.22$ & 6 & .737 \\
External Uncertainty & Context $\times$ evaluation-pathway block & $\chi^2=13.17$ & 6 & .121 \\
Verifiability & Context $\times$ evaluation-pathway block & $\chi^2=3.55$ & 6 & .737 \\
\addlinespace[0.3em]
\multicolumn{5}{l}{\textit{Selected participant-level associations and moderation}} \\
General AI Trust & Main association with Trust (upstream) & $\beta=.202$ & -- & $<.001$ \\
General AI Trust & Main association with Trust (pathway-adjusted) & $\beta=.046$ & -- & .109 \\
General AI Trust & Main association with Reliance (upstream) & $\beta=.227$ & -- & $<.001$ \\
General AI Trust & Main association with Reliance (pathway-adjusted) & $\beta=.053$ & -- & .009 \\
Human Trust Propensity & Safety Concern $\times$ moderator $\rightarrow$ Trust & $\beta_{\mathrm{int}}=.056$ & -- & .002 \\
Objective Financial Literacy & Source Knowledge $\times$ moderator $\rightarrow$ Trust & $\beta_{\mathrm{int}}=.054$ & -- & .017 \\
Objective Financial Literacy & Trust $\times$ moderator $\rightarrow$ Reliance & $\beta_{\mathrm{int}}=.039$ & -- & .002 \\
\bottomrule
\end{tabular}
\vspace{2pt}
\begin{minipage}{0.97\textwidth}
\footnotesize
\textit{Note.} Decision-context entries are likelihood-ratio tests comparing
the additive model with a joint interaction model. For Trust, the interaction
block included Message Appraisal, Risk Acknowledgment, Safety Concern, Source
Knowledge, and Overall Quality; for Reliance, Trust was additionally included.
Benjamini--Hochberg correction was applied within the prespecified test
families. Participant characteristics were standardized in moderation models.
\end{minipage}
\end{table}

\FloatBarrier

\section{Robustness Analyses}
\label{sec:appendix-robustness}

We conducted four sensitivity analyses focused on the central
appraisal--Quality--Trust--Reliance pathway.

\paragraph{Within- and between-participant variation.}
We decomposed each time-varying predictor into a participant mean and a
within-participant deviation and re-estimated the Quality, Trust, and Reliance
models using both components. The principal associations remained at the
within-participant level. Message Appraisal ($\beta_W=.329$), Risk
Acknowledgment ($\beta_W=.073$), Safety Concern ($\beta_W=-.106$), and Source
Knowledge ($\beta_W=.444$) retained their expected associations with Overall
Quality (all $p\leq.001$). Overall Quality remained strongly associated with
Trust ($\beta_W=.573$, $p<.001$), and Trust remained the dominant
within-participant correlate of Reliance ($\beta_W=.533$, $p<.001$).

\paragraph{Alternative repeated-measures estimation.}
Generalized estimating equations with an exchangeable working correlation
closely reproduced the mixed-effects estimates. The Overall Quality--Trust
association was $\beta=.608$ under GEE (vs. $.607$ in the primary model), and
the Trust--Reliance association was $\beta=.595$ (vs. $.594$). The only
notable difference was the small Source Knowledge--Reliance association, which
remained similar in magnitude ($\beta=.048$) but was not statistically
distinguishable from zero ($p=.084$).

\paragraph{Scenario fixed effects.}
Replacing Stakes, External Uncertainty, and Verifiability with fixed effects for
the eight scenarios produced nearly identical estimates. Message Appraisal
predicted Overall Quality at $\beta=.320$ (vs. $.325$), Source Knowledge at
$\beta=.491$ (vs. $.489$), Overall Quality predicted Trust at $\beta=.603$
(vs. $.607$), and Trust predicted Reliance at $\beta=.593$ (vs. $.594$).

\paragraph{Constituent-item sensitivity.}
Replacing Message Appraisal and Safety Concern with their constituent items
preserved the conclusions. The three message-appraisal items were jointly
associated with Overall Quality ($\chi^2(3)=211.04$, $p<.001$), Trust
($\chi^2(3)=35.79$, $p<.001$), and Reliance ($\chi^2(3)=16.67$, $p<.001$).
The two safety-concern items were likewise jointly associated with Overall
Quality ($\chi^2(2)=23.70$, $p<.001$), Trust ($\chi^2(2)=56.55$, $p<.001$),
and Reliance ($\chi^2(2)=12.45$, $p=.002$). The central downstream estimates
also remained nearly unchanged: Overall Quality predicted Trust at
$\beta=.598$ (vs. $.607$), while Overall Quality and Trust predicted Reliance
at $\beta=.208$ and $\beta=.594$ (vs. $.206$ and $.594$, respectively).

\end{document}

\endinput

